\documentclass[lettersize,journal]{IEEEtran}

\usepackage{amsmath,amsfonts}
\usepackage{algorithmic}
\usepackage{array}
\usepackage[caption=false,font=normalsize,labelfont=sf,textfont=sf]{subfig}
\usepackage{textcomp}
\usepackage{stfloats}
\usepackage{url}
\usepackage{verbatim}
\usepackage{graphicx}
\usepackage{balance}
\usepackage{amsmath}
\usepackage{amssymb}
\usepackage{multicol}
\usepackage{multirow}
\usepackage{booktabs}
\usepackage{float}
\usepackage{hyperref}
\usepackage{xcolor}
\usepackage{cite}
\providecommand{\eref}[1]{\eqref{#1}}  
\providecommand{\cref}[1]{Chapter~\ref{#1}}

\providecommand{\R}{\ensuremath{\mathbb{R}}}

\providecommand{\E}{\ensuremath{\mathbb{E}}}

\providecommand{\bydef}{\overset{\text{def}}{=}}

\providecommand{\mat}[1]{\ensuremath{\boldsymbol{#1}}}

\providecommand{\calB}{\mathcal{B}}

\providecommand{\calH}{\mathcal{H}}

\providecommand{\calL}{\mathcal{L}}

\providecommand{\calT}{\mathcal{T}}

\providecommand{\mI}{\mathbf{I}}

\providecommand{\va}{\mathbf{a}}

\providecommand{\vu}{\mathbf{u}}

\providecommand{\vw}{\mathbf{w}}
\providecommand{\vx}{\mathbf{x}}

\providecommand{\vz}{\mathbf{z}}

\providecommand{\mSigma}{\mat{\Sigma}}

\begin{document}
\title{Imaging through the Atmosphere using Turbulence Mitigation Transformer}
\author{Xingguang~Zhang,~\IEEEmembership{Student~Member,~IEEE,}
        Zhiyuan~Mao,
        Nicholas~Chimitt,~\IEEEmembership{Member,~IEEE,}
        Stanley~H.~Chan,~\IEEEmembership{Senior~Member,~IEEE}
\thanks{XZ, NC, and SC are with the School of Electrical and Computer Engineering, Purdue University, West Lafayette,
IN, 47907 USA. Email: \{zhan3275, nchimitt, stanchan\}@purdue.edu. ZM is currently with Samsung Research America. The work was completed while he was at Purdue University, Email: m940421@gmail.com}
\thanks{The research is based upon work supported in part by the Intelligence Advanced Research Projects Activity (IARPA) under Contract No. 2022‐21102100004, and in part by the National Science Foundation under the grants CCSS-2030570 and IIS-2133032. The views and conclusions contained herein are those of the authors and should not be interpreted as necessarily representing the official policies, either expressed or implied, of IARPA or the U.S. Government. The U.S. Government is authorized to reproduce and distribute reprints for governmental purposes, notwithstanding any copyright annotation therein.}}
\maketitle

\begin{abstract}
Restoring images distorted by atmospheric turbulence is a ubiquitous problem in long-range imaging applications. While existing deep-learning-based methods have demonstrated promising results in specific testing conditions, they suffer from three limitations: (1) lack of generalization capability from synthetic training data to real turbulence data; (2) failure to scale, hence causing memory and speed challenges when extending the idea to a large number of frames; (3) lack of a fast and accurate simulator to generate data for training neural networks.

In this paper, we introduce the turbulence mitigation transformer (TMT) that explicitly addresses these issues. TMT brings three contributions: Firstly, TMT explicitly uses turbulence physics by decoupling the turbulence degradation and introducing a multi-scale loss for removing distortion, thus improving effectiveness. Secondly, TMT presents a new attention module along the temporal axis to extract extra features efficiently, thus improving memory and speed. Thirdly, TMT introduces a new simulator based on the Fourier sampler, temporal correlation, and flexible kernel size, thus improving our capability to synthesize better training data.
TMT outperforms state-of-the-art video restoration models, especially in generalizing from synthetic to real turbulence data. Code, videos, and datasets are available at \href{https://xg416.github.io/TMT}{\textcolor{pink}{https://xg416.github.io/TMT}}.
\end{abstract}


\section{Introduction}
\IEEEPARstart{A}{tmospheric} turbulence mitigation methods aim at recovering images distorted by the random fluctuations of the refractive index in the atmosphere. Turbulence-distorted images suffer from spatially varying blurs and geometric warping. These problems, when presented to computer vision applications such as detection, recognition, and surveillance, will cause uncertainty about the object's shape, boundary, and resolution. If they are further entangled with camera shake and sensor noise, restoring images would be challenging. This paper aims to present a new deep-learning method to overcome these challenges.

\begin{figure}[ht]
  \centering   \includegraphics[width=0.99\linewidth]{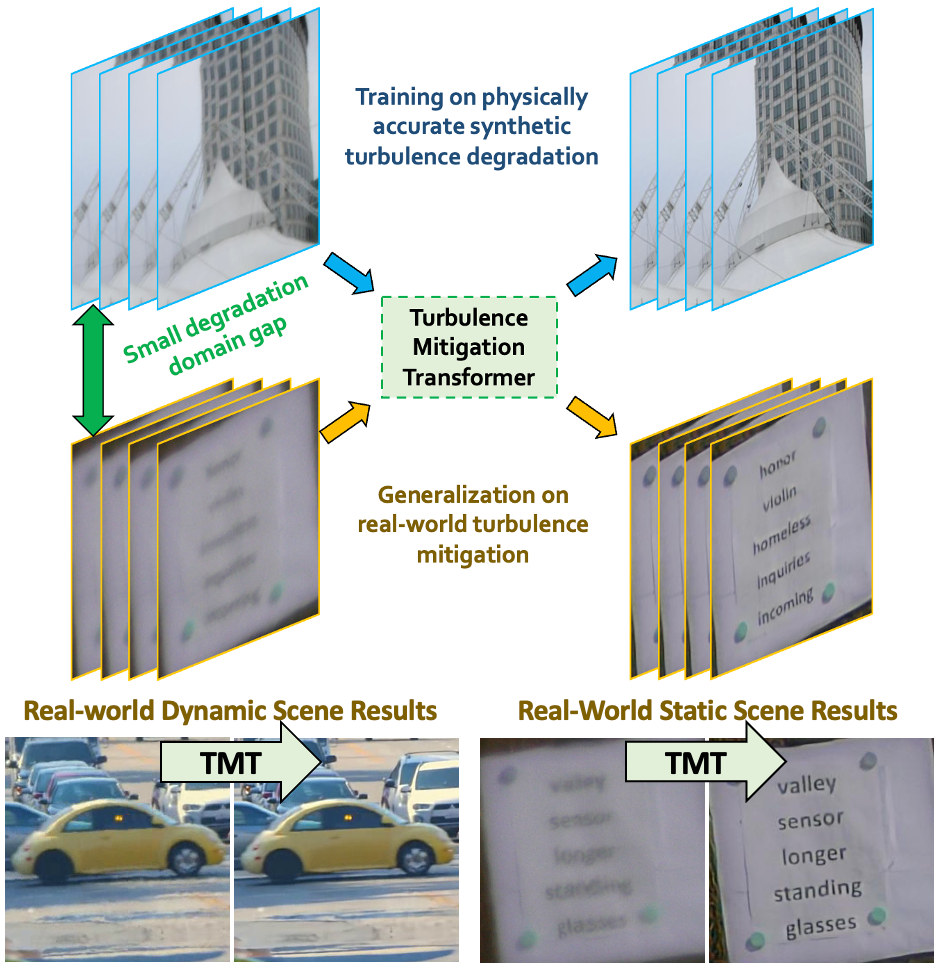}  \caption{The overall idea of our method for turbulence mitigation. Our main contribution includes a physically accurate, diverse, and fast turbulence synthesis and a specifically designed neural network called turbulence mitigation transformer (TMT) for blind multi-frame turbulence removal. The TMT is trained purely on synthetic data in a supervised framework yet shows good generalization and robustness in dealing with real-world turbulence degradation in dynamic and static scenes.}
  \label{fig:intro}
\end{figure}

Image restoration methods for atmospheric turbulence have been studied for decades, yet three gaps remain:
\begin{enumerate}
\item[(1)] Lack of an accurate and fast \emph{simulator} to synthesize training data at a large scale. Existing turbulence simulators (for example \cite{Hardie2017, Lau2021_sim}) are either too slow or inaccurate for the purpose of generating a sufficient amount of training data to train a restoration neural network. The phase-to-space transform (P2S) \cite{mao_P2S} overcomes some difficulties, but it has a severe memory limitation, prohibiting dense-field turbulence simulation.
\item[(2)] Lack of a restoration method that can \emph{generalize} from synthetic data to real turbulence. Classical methods such as \cite{Anantrasirichai2013, mao_tci, lau2019restoration, Milanfar2013, Hirsch2010} are mostly based on optimization using heuristic priors. Their limited modeling capability makes them hard to generalize. Single frame deep-learning methods \cite{Yasarla2021ICIP, Nair2021ICIP, Lau2021} are usually class-specific. Without a strong semantic prior, such as a face, these methods would be vulnerable to out-of-distribution samples.
\item[(3)] Lack of a restoration method that offers both \emph{speed and memory}. Among all the available methods in the literature, multi-frame deep-learning methods (such as \cite{Jin2021NatureMI}) have the greatest potential for the restoration task due to the incorporation of the lucky effect \cite{Fried78}. However, multi-frame methods have a huge memory requirement prohibiting us from using more frames.
\end{enumerate}

In this paper, we present the \textit{Turbulence Mitigation Transformer} (TMT). TMT is the first multi-frame image restoration transformer customized for atmospheric turbulence, with three components articulating the above-mentioned gaps.

\begin{itemize}
\item Recognizing the importance of the forward physical model, TMT explicitly uses the turbulence physics by (1) decoupling the restoration task into de-tilting and de-blurring steps, as opposed to generic video restoration transformers that are single-step networks; (2) introducing a multi-scale structure to supervise the training at different resolutions, as opposed to existing single-scale turbulence networks. These two changes significantly improve the \emph{generalization} capability of TMT.
\item Existing video restoration transformers have severe memory limitations. TMT introduces the concept of temporal-channel joint attention (TCJA), a self-attention module along the temporal axis. TCJA allows TMT to process frames with much less memory, significantly improve the temporal horizon of usable frames, provide better out-sample generalization, and improve \emph{speed}.
\item To support the training of TMT, particularly the multi-scale supervision which requires a highly accurate dense-field simulator, we introduce a new turbulence simulator. The new simulator contains three elements: (1) a Fourier-based sampler to enable dense-field simulation without interpolation; (2) incorporation of the temporal correlation; (3) improvement of the kernel size flexibility.
\end{itemize}

\section{Problem Formulation and Related Work}
\subsection{Turbulence Degradation}
Consider an object $J(\vx,t) \in \R^d$ with a coordinate $\vx \in \R^2$ and time stamp $t \in \R$. As light propagates from the object plane to the image plane, the random fluctuations of the refractive index in the atmosphere will cause turbulence distortion. The exact image formation process remains an open problem, although from Kolmogorov \cite{Kolmogorov1941} to Fried \cite{Fried78}, people have developed widely acceptable models and numerical simulators through wave propagation equations  \cite{Roggemann_1995_a, Hardie2017, SchmidtTurbBook}. For brevity of this paper, we refer readers to tutorials such as \cite{chan2023computational} for discussions of this historical work.

For the purpose of this paper, we shall take an \emph{image processing} perspective (as opposed to the wave propagation perspective) by considering the turbulence as a concatenation of two processes, tilt $\calT$ and blur $\calB$:
\begin{equation}
\underset{\text{corrupted image}}{\underbrace{I(\vx,t)}} = \underset{\text{tilt-then-blur } \calH}{\underbrace{[\calB \circ \calT]}} \quad (\underset{\text{clean image}}{\underbrace{J(\vx,t)}}),
\label{eq: forward model}
\end{equation}
where $\circ$ denotes the functional composition of the two operators, and $\calH \bydef \calB \circ \calT$ is the overall distortion. The foundation of this tilt-blur decomposition is rooted in the Zernike decomposition in the phase space \cite{Roggemann_1995_a}, where it can be shown that the first two Zernike coefficients of the phase are responsible for the pixel displacement, aka \emph{tilt}, and the other Zernike coefficients are responsible for the high order aberrations, aka \emph{blur}. The ordering of the tilt-then-blur does not commute and is justified in \cite{chan_tilt_or_blur}.

\subsection{Existing Turbulence Mitigation Methods}
The problem of turbulence mitigation is to invert the process in \eref{eq: forward model}. For small field-of-view objects such as stars, the turbulence is likely isoplanatic (i.e., spatially invariant), so standard deconvolution applies \cite{primot1990deconvolution}. For larger fields of view, the turbulence becomes anisoplanatic (i.e., spatially varying). Classical methods typically use the lucky selection method \cite{Fried78, vorontsov2001anisoplanatic}, and later a decoupling strategy of de-tilting and de-blurring \cite{shimizu2008super}. During the 2010s, numerous algorithms have been proposed, ranging from numerical optimizations \cite{lou2013video, oreifej2012simultaneous, Milanfar2013}, complex wavelets \cite{Anantrasirichai2013}, Fourier burst accumulation \cite{gilles2016wavelet}, block matching \cite{Hardie2017recon}, and other hybrid methods based on lucky fusion and deconvolution \cite{aubailly2009automated, mao2012non, xie2016removing, lau2019restoration, mao_tci}.

On the deep-learning side, both deterministic \cite{Yasarla2021ICIP, Nair2021ICIP, yasarla2020learning, mao2022single} and generative methods \cite{Lau2021, rai2022removing, nair2023ddpm} have been proposed for single frame restoration. While generative methods or adversarial loss typically produce visually better images, they could also be more vulnerable to small perturbations on input \cite{Choi_2019_ICCV, choi2022deep}. Among the above methods, similar to our work, \cite{Lau2021, yasarla2020learning} consider removing tilt and blur progressively in two stages, while others mitigate turbulence degradation in one stage. \cite{feng2023turbugan, Li_2021_ICCV} introduced unsupervised methods to restore a single image from multiple instances. Although the large-scale dataset is not required, these works need several minutes or hours to converge and can only be applied to static scene sequences. \cite{Jin2021NatureMI} is the only work for general multi-frame turbulence mitigation in the literature; it is faster and covers dynamic scenes, but its generalization capability is limited.

\subsection{Learning-based Image and Video Restoration Methods}
Arguably, solving \eref{eq: forward model} shares many similarities with a video restoration problem. Classical algorithms have been used directly for turbulence \cite{chan_tv_video}. Therefore, when moving to deep learning, it is fair to expect that existing state-of-the-art video restoration networks such as \cite{liang2022vrt, chan2022basicvsrpp} can be re-trained with appropriate data to perform the task. 

Since turbulence degradation is spatially and temporally variant, the self-attention mechanism in vision transformers is well suited because of their adaptive weights based on the semantics of the image, while convolution-based networks use the same kernel on the entire image and feature map. Additionally, in single-frame restoration, transformers such as \cite{Wang2022Uformer, liang2021swinir, zamir2021restormer} have been introduced and shown superior performance. 
However, when developing transformer-based models for multi-frame turbulence mitigation, the memory requirement is the biggest bottleneck for restoration methods. While extending the idea to the temporal dimension is possible \cite{liang2022vrt}, the computation and memory will make the implementation impractical. One of the key contributions of TMT is a new module, TCJA, to overcome this memory problem.

\subsection{Existing Turbulence Simulators}
When using deep learning based restoration methods, data is of the utmost importance. Since collecting real turbulence for \emph{training} is nearly impossible due to the scale and variety of weather conditions we need, simulation is the best alternative. Turbulence simulation tools are mostly developed for physics and defense applications where precision is the most critical consideration. Split-step \cite{Hardie_2017_a, Roggemann_2012_a, Roggemann_1995_a, Schmidt_2010_a} is often regarded as the ``gold standard'' although it is extremely slow. While faster simulations exist \cite{Milanfar2013, Lau2021_sim}, their accuracy is inadequate for our purpose. The simulator we present here is an improvement of the phase-to-space transform reported in \cite{chimitt2020simulating, mao_P2S}. For detailed discussions of the latest development of turbulence simulations, we refer the readers to \cite{chan2023computational}.

\section{Turbulence Mitigation Transformer}
\subsection{Design Philosophy}
To motivate the design of TMT, we start by looking at how vision transformers today are used in deconvolution problems. Vision transformers use self-attention to direct the neurons to focus on spatially correlated features \cite{dosovitskiy2020image,liu2021Swin}. Transformers are particularly adaptive to \emph{spatially varying} distortions because the self-attention can effectively construct \emph{localized} kernels instead of a global kernel in a convolutional neural network \cite{mao2022single}.

When designing TMT, we need to revisit the forward model outlined in \eref{eq: forward model}. The forward model says that the turbulence distortion is \emph{compositional} --- it contains tilt and blur. The ways to handle these two types of distortions are different:
\begin{itemize}
\item \textbf{Tilt}: Tilt is a temporal problem. Turbulent tilts follow a zero-mean
Gaussian process \cite{Fried78}. Over a long period of \emph{time}, pixels will experience the lucky phenomenon which is the foundation of many classical turbulence methods \cite{aubailly2009automated, mao2012non, xie2016removing, lau2019restoration, mao_tci, Milanfar2013}. \cite{Li_2021_ICCV} shows that a small network is sufficient to rectify the images by measuring the distortion field and sampling pixels on its implicit ```mean position". To incorporate this idea, TMT introduces a \emph{lightweight depth-wise 3D UNet}.
\item \textbf{Blur}: Blur is a spatial \emph{and} temporal problem where high-order aberrations in the phase cause the image to suffer from non-uniform blurs across the field of view. Spatial-temporal transformers are not easy to implement because of the huge memory consumption. To overcome the memory issue, TMT introduces a new temporal-channel joint attention (TCJA) module.
\end{itemize}

\begin{figure*}[ht]
  \centering
   \includegraphics[width=\linewidth]{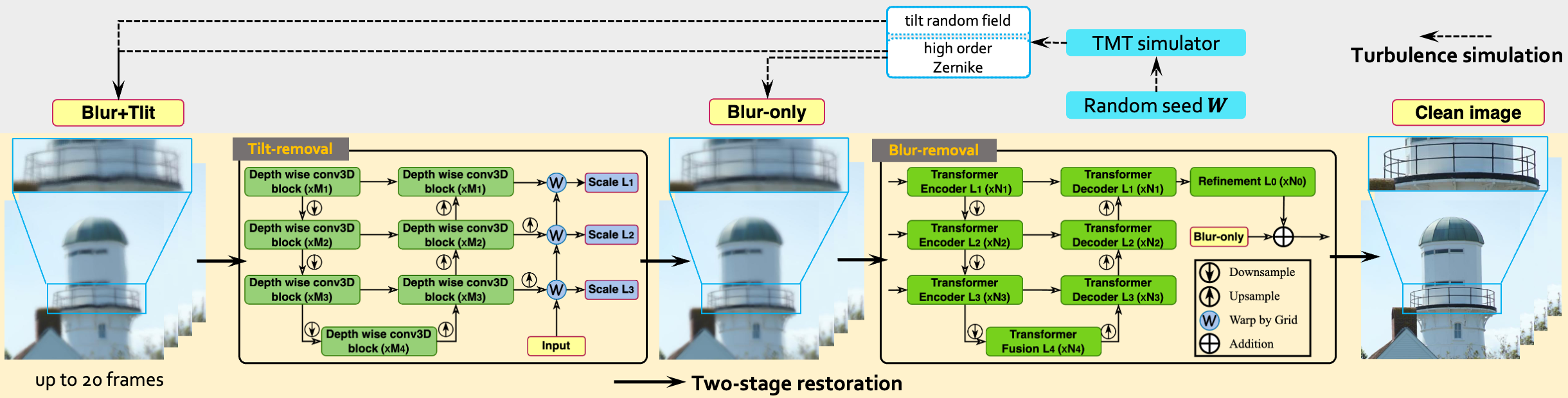}
  \caption{Illustration of the simulation and restoration pipeline. The two-stage design of the restoration can take advantage of the physics-based simulation method. The lighthouse image is from the training set of our dataset.}
  \label{fig:pipeline}
\end{figure*}

To summarize, TMT is a customized transformer-based restoration for \emph{turbulence}. We summarize its key properties in Table~\ref{tab: transformer} and its structure in Fig. \ref{fig:pipeline}.

\begin{table}[t]
\caption{Comparison between existing SOTA video restoration and turbulence mitigation methods and the proposed TMT.}
\renewcommand{\arraystretch}{1.2}
\begin{tabular}{p{4cm}p{4cm}}
\hline
Existing video restoration and turbulence mitigation networks \cite{liang2022vrt, Jin2021NatureMI, chan2022basicvsrpp} & TMT (Proposed) \\
\hline
$\bullet$ Single-stage  & $\bullet$ Two-stage: tilt + blur \\
$\bullet$ Agnostic to turbulence    & $\bullet$ Customized for turbulence\\
$\bullet$ Local spatial attention     & $\bullet$ Temporal-channel attention\\
$\bullet$ Temporal modeling: N.A. / local window based / recurrent & $\bullet$ Temporal modeling: Fully connected \\
$\bullet$ Single-scale input        & $\bullet$ Multi-scale loss (for tilt) and input (for blur)\\
\hline
\end{tabular}
\label{tab: transformer}
\vspace{-2ex}
\end{table}

\subsection{Tilt-Removal Module}
The tilt-removal module aims to align the images to the greatest extent. Since a turbulence scene contains both moving objects and pixel jittering, a direct temporal fusion (as in naively extending the spatial transformer to a spatial-temporal transformer) cannot extract features consistently without demanding a large memory consumption.

Our proposed tilt-removal module is based on a depth-wise 3D convolution. We chose this structure for two reasons:
\begin{itemize}
\setlength\itemsep{0ex}
\item While classical video restoration methods (in the absence of turbulence) use optical flow \cite{ranjan2017optical} or deformable convolution \cite{dai2017deformable, chan2022basicvsrpp} to align the object motion, they are limited to \emph{two adjacent frames}. Our 3D convolution does not have this limitation. As such, we can align multiple frames simultaneously.
\item Recent works \cite{Li_2021_ICCV, anantrasirichai2022atmospheric} have shown that a complicated network is not needed for tilt-removal. A simple network is often enough to remove most pixel jitters due to the random tilt. Thus, we chose lightweight depth-wise 3D convolution inspired by these previous works.
\end{itemize}

The tilt-removal module is summarized in Table~\ref{tab: T-remover}. The basic structure is a UNet. We use a depth-wise (DW) 3D convolution and ReLU. It has four levels. Following the last three levels of the decoder, we output three sequences of warping grids to rectify the input image progressively. The warping grids from the 3rd and 2nd levels are upsampled to have the same spatial dimension as the input. To enlarge the perceptual field of the network, we set the kernel sizes of the first three encoder levels as 7, 7, and 5.

\begin{table}[h]
\caption{Tilt-removal module.}
  \centering
\resizebox{0.48\textwidth}{!}{
\begin{tabular}{c|c|c}
\hline
Aspects & Encoder Level [1,2,3,4] & Decoder level [3,2,1] \\
\hline
Type of convolution & [Conv, DW, DW, DW] & [DW, DW, DW] \\
Size of kernels    &  [7, 7, 5, 3] & [3, 3, 3]\\
$\#$ output channels  &  [64, 128, 256, 512] & [256, 128, 64]\\
Down/Up-sampling  &  Max-pooling & Transposed Conv. \\
\hline
\end{tabular}}
\label{tab: T-remover}
\end{table}

\subsection{Multi-Scale Loss}
A critical component of TMT's tilt-removal module is the multi-scale loss. The multi-scale loss is also how we inject physics into the restoration model to improve generalization.

Assuming we have a faithful simulator, we will be able to generate \emph{tilt-free} blur-only images at \emph{multiple scales}:
\begin{equation}
\underset{\text{tilt-free at scale $\ell$}}{\underbrace{\widetilde{J}_{\ell}(\vx,t)}} = \underset{\text{blur at scale $\ell$}}{\underbrace{\calB_{\ell}}} ( \; \underset{\text{ground truth}}{\underbrace{J(\vx,t)}} \; ), \quad \ell = 1,\ldots,L,
\label{eq: blur only}
\end{equation}
where $L = 3$ is the number of scales. The difference between \eref{eq: blur only} and \eref{eq: forward model} is that there is no tilt in \eref{eq: blur only} but only blur. Therefore, we are preparing tilt+blur and blur-only pairs to train the tilt-removal module.

The tilt-removal module outlined in the previous subsection will return us, at every scale, an estimate of the tilt-free image:
\begin{equation}
\underset{\text{estimated tilt-free at scale $\ell$}}{\underbrace{\widehat{J}_{\ell}(\vx,t)}} = \underset{\text{tilt-removal}}{\underbrace{\calT_{\ell}^{-1}}} ( \; \underset{\text{fully distorted}}{\underbrace{ \{I(\vx,t)\}}} \; ),
\end{equation}
where $\calT_{\ell}^{-1}$ is a symbolic representation of the branch in the tilt-removal module that generates an image at scale $\ell$. The input to $\calT_{\ell}^{-1}$ is a stack of $T$ frames $\{I(\vx,t)\}$ instead of a single frame at time $t$.

Once the tilt-free ground truths and the tilt-free estimates are determined, we can define the loss function as
\begin{equation}
\calL_{\text{tilt}} = \sum_{\ell=1}^L \;\; \underset{\text{weight}}{\underbrace{\gamma_\ell}} \; \cdot \underset{\text{Charbonnier loss}}{\underbrace{\calL_{\text{char}}\left( \widehat{J}_{\ell}(\vx,t), \widetilde{J}_{\ell}(\vx,t) \right)}} ,
\end{equation}
where $\gamma_{\ell}$ denotes the weight and $\calL_{\text{char}}$ is the Charbonnier loss \cite{charbonnier}. The choice of the weight is determined empirically as $\gamma_{1} = 0.6$, $\gamma_{1} = 0.3$ and $\gamma_{1} = 0.1$

The intuition of our multi-scale loss is that while tilts can be highly random at full resolution, they are much more structured and weaker at lower resolution. Therefore, tilt-removal is often easier at the lower scales and gradually becomes harder as we progress the scales. As a result, by \emph{progressively} minimizing the loss across scales, we ensure that the tilt-removal is scale-consistent.

\begin{figure*}[t]
  \centering
   \includegraphics[width=0.9\linewidth]{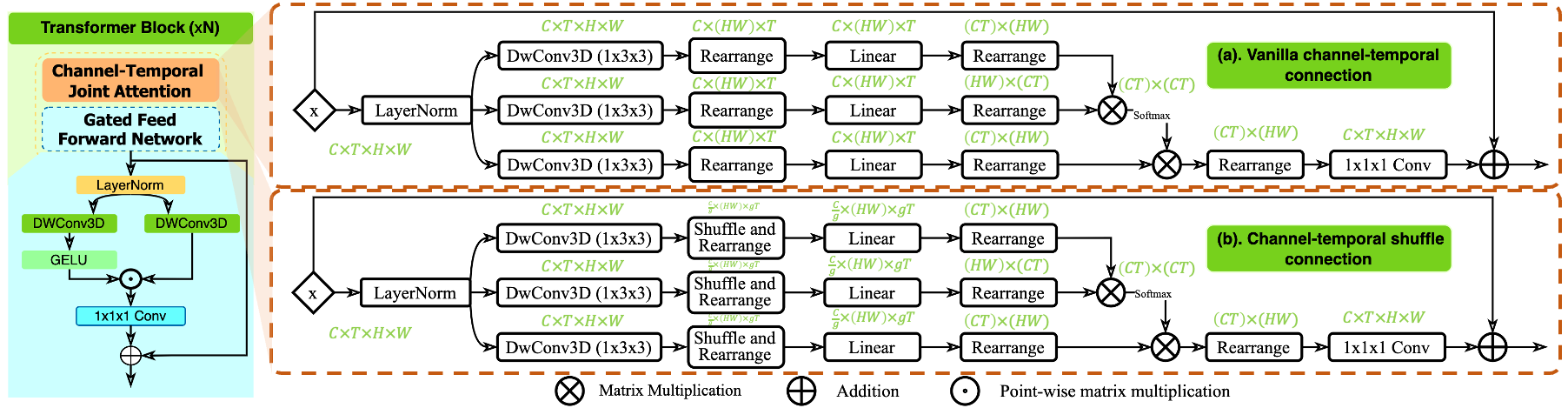}
  \caption{The architecture of the transformer block. It consists of the channel-temporal joint attention (CTJA) module followed by a gated feed-forward network. As shown in the dashed box, we designed two variants (a) and (b) of the CTJA block for $\text{TMT}_{a}$ and $\text{TMT}_{b}$, respectively. The only difference between those two is the channel shuffle operation.}
  \label{fig:attention}
\end{figure*}

\subsection{Blur-Removal Module}
The challenge of designing the blur-removal module is how to construct global temporal attention of all the frames without suffering from a high complexity and memory requirement. In conventional transformers, generating attention has a complexity that grows quadratically with the number of pixels in the video. Thus, conventional transformers adopt a window-based strategy to compute the attention locally \cite{liang2022vrt}. 

TMT's deblurring module is shown on the right-hand side of the restoration pipeline in Fig. \ref{fig:pipeline}. The overall structure is similar to the tilt-removal module, except that basic operations are defined through a new transformer block outlined in Fig. \ref{fig:attention}. The key merits of this new transformer block are
\begin{enumerate}
\item In TMT's transformer, the spatial coordinates are connected purely via convolution layers. This smooths the interaction among neighboring pixels, avoiding inconsistent performance around the margin of local windows.
\item The core module of TMT is temporal-channel self-attention (TCJA). We compute the self-attention matrix for each pixel on the temporal and channel axes. Although the convolution operation is spatially invariant, the combination of the features of each pixel is spatially independent, enabling spatially varying restoration.
\item The dynamics of the blur and residue jitter largely follow a zero-mean random process. Therefore, by enabling full temporal connection over more frames, TMT can be more efficient than conventional transformers in capturing temporal dynamics.
\end{enumerate}

On the right-hand side of Fig. \ref{fig:attention}, we present two variants of the proposed temporal channel joint attention (TCJA). As TCJA constructs attention along the temporal and channel dimensions, a standard way for self-attention should be flattening these two dimensions first, then using a linear projection to map it into the key, query, and value spaces. However, as we use a large chunk of frames (12 $\sim$ 20 frames), the temporal-channel vector is large, and the projection will require $O(\frac{HWC^2T^2}{h})$ complexity, where $h$ is the number of the head, $H$, $W$, $T$ are height, width, and the number of frames. We propose two low-rank projection strategies to overcome this, as summarized below:
\begin{itemize}
\item In (a), we separate the temporal and channel axes and apply a linear projection on the temporal dimension. In this case, features in different channels of different frames do not communicate across frames before the self-attention step. This offers a slight boost in speed. 
\item In (b), we implement channel shuffle and partial connection inspired by the ShuffleNet \cite{zhang2018shufflenet}. We split the channels into groups and connect the temporal axis with the channels in each group, allowing them to communicate across several attention blocks.
\end{itemize}

\section{Improvement for Simulator}
Like any deep learning algorithm, the success and failure of TMT are intimately related to how good our data is and how much data we have. For TMT, the source of \emph{training} data is synthesized by our turbulence simulator modified from \cite{chimitt2020simulating, mao_P2S, chimitt2022real}, with several important improvements.

\subsection{Basic Principles}
The starting point of our simulator is Zernike-space simulation \cite{chimitt2022real}, also known as
the phase-over-aperture model \cite{chimitt2020simulating}. For an input $J(\vx,t)$ at coordinate $\vx$ and time $t$, we directly generate the spatially varying point spread function (PSF) and compute the output $I(\vx,t)$ via the equation
\begin{equation}
I(\vx,t) = \int \underset{\text{PSF from $\vu$ to $\vx$ at time $t$}}{\underbrace{h(\vx,\vu,t)}} J(\vu,t) \; d\vu,
\label{eq: simulation 1}
\end{equation}
where $\vx$ denotes the coordinate in the output space (i.e., the image plane) and the $\vu$ denotes the coordinate in the input space (i.e., the object plane). The PSF can, in theory, be obtained through
\begin{equation}
h(\vx,\vu,t) = |\text{Fourier}(e^{j\phi_{\vx,t}(\vu)})|^2
\end{equation}
where $\phi_{\vx,t}$ is the phase function at $(\vx,t)$. Represented in the Zernike space, we can write $\phi_{\vx,t}(\vu) = \sum_{m=1}^M a_{\vx,t}(m) Z_m(\vu)$ with $Z_m(\vu)$ being the $m$-th Zernike basis function, and $a_{\vx,t}(m)$ is the corresponding Zernike coefficient \cite{chimitt2020simulating}.

The computational bottleneck in \eref{eq: simulation 1} is the generation of the Zernike coefficients $a_{\vx,t}(m)$, because once the Zernike coefficients are known, the PSF can be efficiently implemented through a basis decomposition as proposed in \cite{mao_P2S}:
\begin{equation}
h(\vx,\vu,t) = \sum_{k=1}^K \beta_{\vx,t}(k) \varphi_{k}(\vu).
\label{eq: simulation 2}
\end{equation}
In \eref{eq: simulation 2}, the basis function $\varphi_{k}(\vu)$ can be determined through a principal component analysis (or a pre-defined set of basis functions). The basis coefficients $\beta_{\vx,t}(k)$ are computed through the phase-to-space transform (P2S) \cite{mao_P2S} where we use a shallow network to map the Zernike coefficients:
\begin{equation}
\{a_{\vx,t}(m)\} \;\; \underset{\text{P2S network}}{\longrightarrow} \;\; \{\beta_{\vx,t}(k)\}.
\label{eq: simulation 3}
\end{equation}
Here, the bracket-to-bracket mapping means that we are transforming the entire set of Zernike coefficients $\{a_{\vx,t}(m)\}$ to the set of basis coefficients $\{\beta_{\vx,t}(k)\}$.

The DF-P2S simulator \cite{chimitt2022real} resolves another bottleneck issue, which is memory consumption. For the P2S transform in \eref{eq: simulation 3}, we need to first generate the Zernike coefficients $\{a_{\vx,t}(m)\}$. However, generating these Zernike coefficients requires us to construct a 6-dimensional Gaussian covariance tensor (two dimensions for space and one for Zernike order, then duplicate to create the correlation). This has not yet included the time axis, which will add another two dimensions. Since this covariance matrix (technically a tensor) is extremely large, P2S is limited to generating the Zernike coefficients at a grid of points and interpolating otherwise\footnote{The interpolation of phase-to-space is done in the Zernike space. This should be distinguished from interpolating the PSF in \emph{space}, such as \cite{chimitt2020simulating}, \cite{Hardie2017}.}.

In \cite{chimitt2022real}, the memory bottleneck is mitigated by a few approximations to retain the homogeneity (aka wide-sense stationarity) of the covariance matrix. Doing so allows us to use Fast Fourier Transform to draw samples from the covariance matrix. As a result, we do not have to store the covariance matrix because homogeneity allows us to use one slice of the covariance matrix. In terms of speed, we have improved the speed of the covariance matrix generation from 4 minutes for a $512\times 512$ image to 17ms.

\subsection{Add Temporal Correlation}
The first improvement is the addition of the temporal correlation. Recalling the phase-to-space equation in \eref{eq: simulation 3}, the transformation from left to right requires us to start with the Zernike coefficients $\{a_{\vx,t}(m)\}$. For any fixed $(\vx,t)$, we can define a Zernike coefficient vector $\va_{\vx,t} = [a_{\vx,t}(1),\ldots,a_{\vx,t}(M)]^T$ where $M = 36$ is the typical number of coefficients we use. Enumerating over all the possible coordinates $\vx_1,\vx_2,\ldots,\vx_N$, we can construct a very long vector $\overline{\va}_t = [\va_{\vx_1,t}; \, \va_{\vx_2,t}; \, \ldots ;\, \va_{\vx_N,t}]$ where $N$ denotes the number of pixels of the image. Generation of $\overline{\va}_{t}$ is done by multiplying a white noise vector $\vw_t \overset{\text{i.i.d.}}{\sim} \text{Gaussian}(\mathbf{0},\mI)$ with a huge covariance matrix $\mSigma$:
\begin{equation}
\overline{\va}_{t} = \mSigma^{\frac{1}{2}} \vw_t, \qquad t = 1,2,\ldots,T
\label{eq: AR 1}
\end{equation}
where $\mSigma^{\frac{1}{2}}$ accounts for the spectral factorization of $\mSigma$. Using the sequence of approximation in \cite{chimitt2022real}, the spectral factorization can be done using the Fast Fourier Transform.

The temporal correlation of adjacent frames of a turbulence video can theoretically be modeled via Taylor's frozen hypothesis \cite{roggemann1996imaging}. This would expand our 6-dimensional Zernike tensor to 8 dimensions. This is certainly doable in theory, but it is just computationally infeasible. To alleviate this computational difficulty, we adopt a surrogate approach by considering the auto-regressive model, where we define
\begin{equation}
\vw_t = \alpha \vw_{t-1} + \sqrt{1-\alpha^2} \vz,
\label{eq: AR 2}
\end{equation}
where $\vz \overset{\text{i.i.d.}}{\sim} \text{Gaussian}(\mathbf{0},\mI)$ and $0 \leq \alpha \leq 1$. It is easy to show that $\E[\vw_t] = \mathbf{0}$ because $\vz$ is zero-mean, and $\E(\vw_t \vw_t^T) = \mI$. Furthermore, we can show that
\begin{equation}
    \E[\vw_t \vw_{t - \tau}^T] = \alpha^\tau \mI
\end{equation}
where $\tau$ is an integer. Thus, the correlation between frames decays as the frames are farther apart.

We remark that in TMT's simulator, the temporal correlation is realized through the \emph{random seed} instead of the final samples. The impact on the final samples can be seen by substituting \eref{eq: AR 2} into \eref{eq: AR 1}
\begin{align*}
\overline{\va}_{t}
&= \mSigma^{\frac{1}{2}} \left(\alpha \vw_{t-1} + \sqrt{1-\alpha^2} \vz\right) \\
&= \alpha \overline{\va}_{t-1}  + \sqrt{1-\alpha^2} \mSigma^{\frac{1}{2}} \vz.
\end{align*}
In other words, we introduce a recurrent relationship by regressing the current samples with the previous samples. The regression parameter $\alpha$ is typically set as $\alpha = 0.9$ for strong correlations and smaller values for weaker correlations.

\subsection{Kernel Size}
Although the DF-P2S simulator \cite{chimitt2022real} offers a significant improvement over the P2S version \cite{mao_P2S}, both assume a \emph{fixed} support (aka blur kernel size) of the PSF $h(\vx,\vu,t)$. More specifically, the kernel size is always $33 \times 33$, regardless of the system parameters. This is clearly a mismatch to reality, where the kernel size depends on the system.

To enable a wide range of kernel sizes, we have re-examined the PSF basis functions of \cite{mao_P2S}. We have observed that the basis functions $\varphi_k$ from the P2S model may incur aliasing due to incorrect interpolation by the resizing process. To address this, we have regenerated $\varphi_k$ at a higher resolution with more training examples. The PCA is, therefore, more representative of the distribution and contains additional high-frequency information. The result is a set of basis functions that may be resized without introducing any noticeable amount of aliasing into the PSF representation.

TMT's simulator includes kernel sizes from $9 \times 9$ to $33 \times 33$. The choice of kernel size is randomly picked to ensure sufficient coverage of the realistic turbulence conditions. In real-world static scene data such as \cite{gilles2017open, UG2}, the field of view is usually very narrow, and the corresponding blur kernels are usually large. In real-world dynamic scene data such as \cite{anantrasirichai2022atmospheric, Jin2021NatureMI, gilles2017open}, turbulence conditions have more variety and most samples have a larger field of view with smaller blur kernels. Overall, we selected the kernel sizes based on all publicly available real-world datasets, and our experiments show that our synthetic data facilitates good generalization.

\begin{table*}[t]
  \caption{Parameter sampling setting of the data synthesis. $U(a,b)$ denotes uniform distribution in the range of $(a,b)$.}
  \label{table:parameter}
  \centering
\setlength{\aboverulesep}{0.2pt}
\setlength{\belowrulesep}{0.2pt}
  \resizebox{\textwidth}{!}{
  \begin{tabular}{ccccccc}
    \toprule[1pt]
    Dataset    & Proportion & Kernel size
    & Aperture (m) & $D/r_{0}$ & Distance (m) & Temporal correlation \\
    \midrule
    \multirow{3}{*}{Static}
       & 20\% (weak) & [33]  & $U(0.001, 0.005)$ & [0.5, 1, 1.2, 1.5] & $U(100, 400)$ & $U(0.2, 0.6)$\\
       & 40\% (medium) & [33]  & $U(0.04, 0.1)$ & [1, 1.5, 2] & $U(400, 800)$ & $U(0.2, 0.6)$\\
       & 40\% (strong) &  [33] & $U(0.1, 0.2)$ & [1.5, 2, 3] & $U(800, 1500)$ & $U(0.2, 0.6)$\\
    \midrule
    \multirow{3}{*}{Dynamic}
     & 33\% (weak) & [9, 13, 15, 21]   & $U(0.001, 0.005)$ & [0.3, 0.6, 1, 1.2] & $U(50, 400)$ & $U(0.4, 0.8)$ \\
     & 33\% (medium) & [11, 17, 25, 33]  & $U(0.04, 0.1)$ & [0.3, 1, 1.5] & $U(400, 800)$ & $U(0.8, 0.95)$\\
     & 33\% (strong) & [15, 21, 27, 33]  & $U(0.1, 0.2)$ & [1, 1.5, 2, 2.5] & $U(800, 2000)$ & $U(0.88, 0.95)$ \\
    \bottomrule[1pt]
  \end{tabular}}
\end{table*}

\subsection{TMT Dataset}
A critical missing piece in the turbulence restoration literature is training data. The TMT simulator is able to simulate the data, but the source images and parameter space controlling the turbulence profile need to be clarified. The engineering question here is then \emph{what} kind of scenarios should we simulate, and \emph{how much} turbulence should we inject? Ultimately, can we create a training dataset for the community to use instead of re-running our simulator?

To this end, we introduce the TMT dataset, which consists of static and dynamic parts. We use the place dataset \cite{zhou2017places} for synthesizing static scenes. We randomly selected 9,017 images in the original dataset as input for the simulator. We generated 50 turbulence images and their associated distortion-free images for every single input, resulting in 9,017 pairs of image sequences of static scenes. We split them into 7,499 pairs and 1,518 pairs for training and testing, respectively.

For dynamic scenes, the TMT dataset contains many videos. The source datasets for our dynamic scene data are the Sports Video in the Wild (SVW) dataset \cite{safdarnejad2015sports} and all ground truth videos used in TSRWGAN \cite{Jin2021NatureMI}. These videos are mixed, generating 4,684 samples with a total number of frames of 1,979,564. We generated 4,684 pairs of full turbulence and distortion-free videos, then randomly split them into 3,500 videos for training and 1,184 for testing, keeping at most 120 frames per testing video.

For synthesizing turbulence data, we have identified key turbulence parameters shown in Table~\ref{table:parameter}. We partition turbulence parameters in three levels: weak, medium, and strong. From experience, we notice that a long distance requires a larger diameter of the aperture. Also, a smaller Fried parameter \cite{Fried78} implies stronger turbulence, which further requires a larger blur kernel to produce. We empirically set temporal correlations to match the visual perception of the existing real-world data for better generalizability.

Table~\ref{tab: TMT dataset} summarizes our dataset. The TMT dataset is by far the largest and most comprehensive dataset for atmospheric turbulence deep learning research. Our dataset is open to the public at \href{https://xg416.github.io/TMT}{\textcolor{pink}{https://xg416.github.io/TMT}}.

\begin{table}[t]
\caption{Specification of the TMT dataset, where each sequence for the static scene data has 50 frames.}
\begin{tabular}{p{1.5cm}p{2cm}p{4cm}}
\hline
& Static & Dynamic \\
\hline
Source & Place \cite{zhou2017places} & Sports \cite{safdarnejad2015sports} and TSRWGAN \cite{Jin2021NatureMI}\\
Amount & 9,017 sequences & 4,684 videos (1,979,564 frames)\\
Training & 7,499 sequences & 3,500 videos \\
Testing  & 1,518 sequences & 1,184 videos \\
\hline
\end{tabular}
\label{tab: TMT dataset}
\end{table}

\section{Experiments}

\subsection{Testing Data}
Our testing data consists of two parts. For quantitative evaluation (which involves PSNR calculations), we use the testing portion of the TMT dataset. We remark that while the testing data is technically in-distribution, they are generated independently from the training data. Ground truth images used for testing are never seen during training.

To examine the generalization of TMT, we test TMT and all competing methods using data from different sources, including the static patterns from OTIS \cite{gilles2017open} dataset, dynamic scene videos from \cite{Jin2021NatureMI} and the CLEAR \cite{anantrasirichai2022atmospheric} dataset. These datasets together contain a wide range of turbulence conditions. We present samples testing results on each in Fig. \ref{fig:OTIS}, \ref{fig:TSRWGAN_Data}, and \ref{fig:WGAN_comparison}.

\subsection{Training Pipeline}
We compare the proposed TMT model with the TSRWGAN \cite{Jin2021NatureMI}, a recent multi-frame turbulence mitigation method and two state-of-the-art general video restoration methods VRT \cite{liang2022vrt}, and BasicVSR++ \cite{chan2022basicvsrpp, chan2022generalization}.

All models are trained separately for static and dynamic scene datasets. We trained our model using the Adam optimizer with the Cosine Annealing scheduler \cite{loshchilov2016sgdr}. The VRT, BasicVSR++, and TSRWGAN models are all trained under the same settings as the original paper. For VRT and BasicVSR++, we use their configurations for motion deblurring as it most closely resembles our task. For a fair comparison, we set the input and output as 12 frames for TMT, VRT, and BasicVSR++. We fine-tuned the pre-trained TSRWGAN model for 600K iterations on our dataset to maximize the performance. Hence, we kept the original 15 input frames setting, which should not put the method at any disadvantage. For the BasicVSR++ \cite{chan2022basicvsrpp}, we trained it from scratch for 600K iterations with batch size 4. The other training scheme is the same as the original paper.

We first trained the tilt-removal module for 400K iterations, and the batch size was set to 4. We used Adam optimizer \cite{kingma2014adam} and the learning rate was initialized as $2\times10^{-4}$ then gradually decreased to $1\times 10^{-6}$ by the Cosine Annealing schedualer \cite{loshchilov2016sgdr}. We fixed the tilt-removal module later and trained the deblurring module for 600K iterations. The same batch size, optimizer, and learning rate scheduler were applied.

For all experiments, we used the same data augmentation strategy: random Gamma correction, random saturation, and random cropping, followed by a random combination of horizontal/vertical flipping and $90^{\circ}$ rotation. Since real turbulence images are often captured using telephoto lenses with large F-number, a high ISO is required to improve the image quality, inevitably leading to non-negligible image noise. Therefore, we also injected Gaussian random noise from $ \mathcal{N}(0, \beta \boldsymbol{I})$, where $\beta \sim \text{uniform}(0,0.0004)$ during training to enhance the robustness of the models.

\begin{figure}
    \captionsetup[subfloat]{font=scriptsize,farskip=1pt}
    \centering
  \subfloat[Input\label{fig2:Input}]{%
       \includegraphics[width=0.49\linewidth]{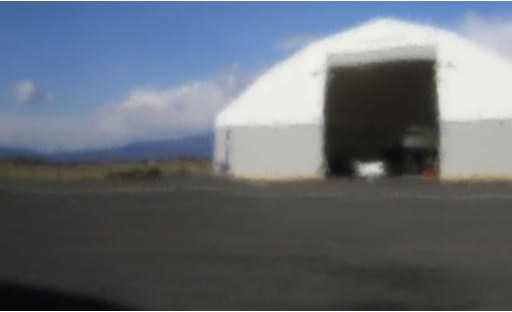}}
    \hfill
  \subfloat[Ground truth\label{1b}]{%
        \includegraphics[width=0.49\linewidth]{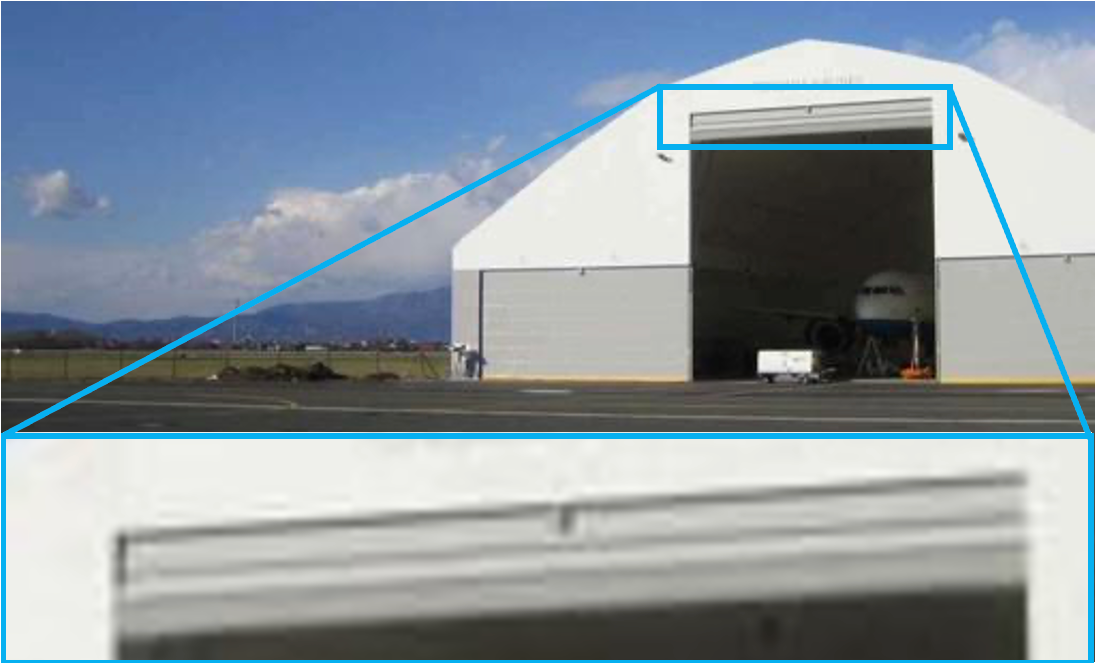}}
    \\
  \subfloat[BasicVSR++ \cite{chan2022basicvsrpp} \label{1c}]{%
        \includegraphics[width=0.49\linewidth]{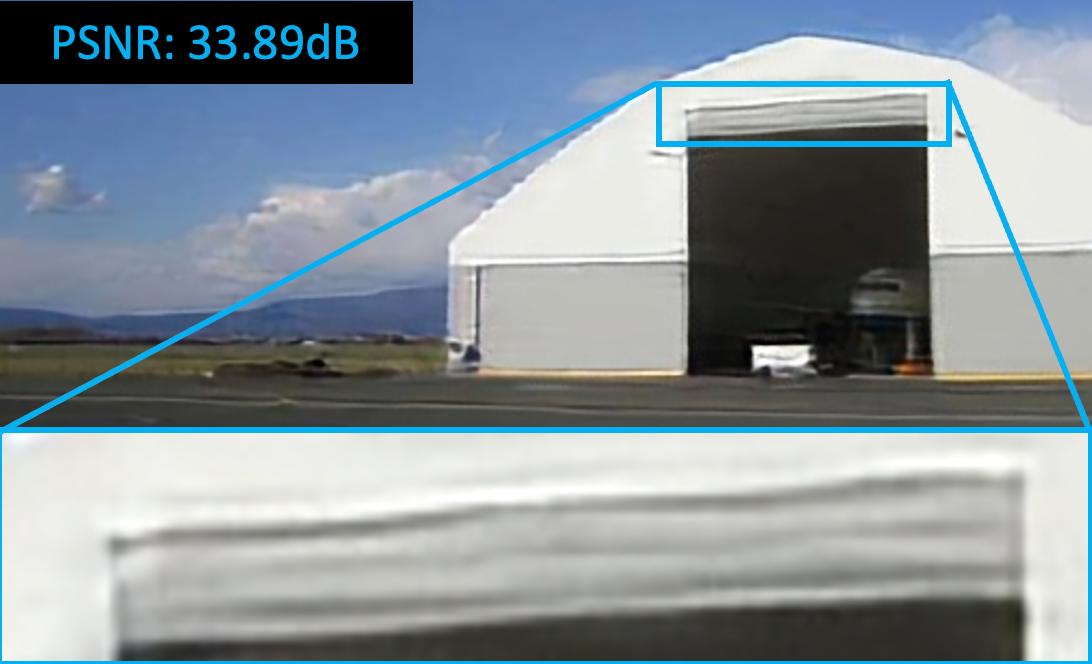}}
  \hfill
  \subfloat[TSRWGAN \cite{Jin2021NatureMI}]{%
        \includegraphics[width=0.49\linewidth]{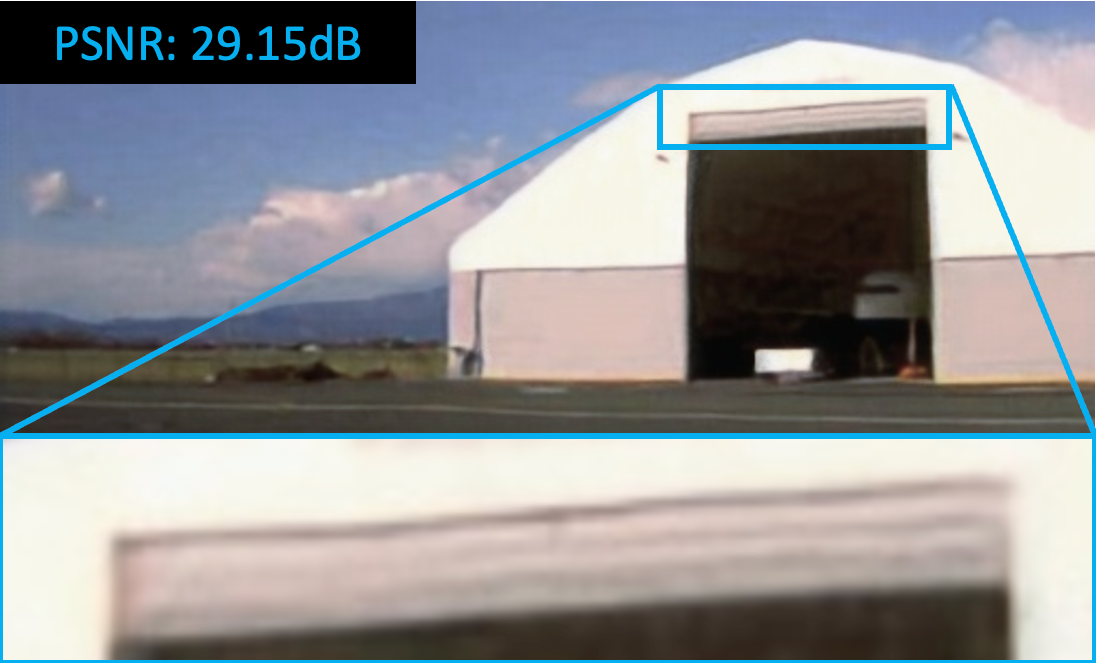}}
    \\
  \subfloat[VRT \cite{liang2022vrt}\label{1e}]{%
        \includegraphics[width=0.49\linewidth]{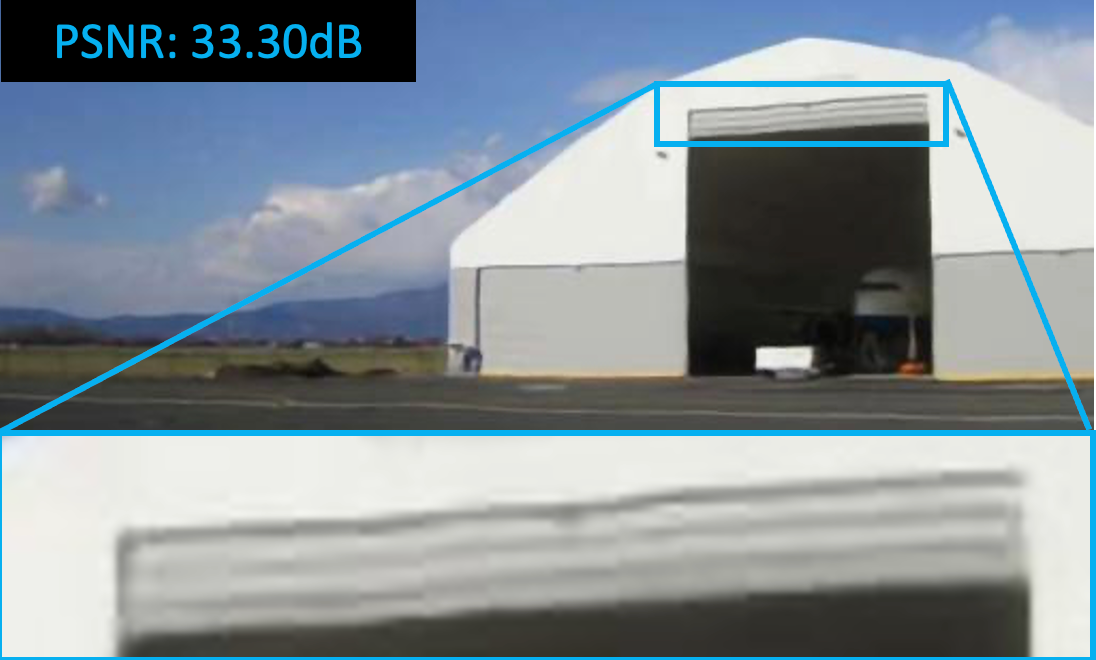}}
    \hfill
  \subfloat[{TMT [ours]} \label{1f}]{%
        \includegraphics[width=0.49\linewidth]{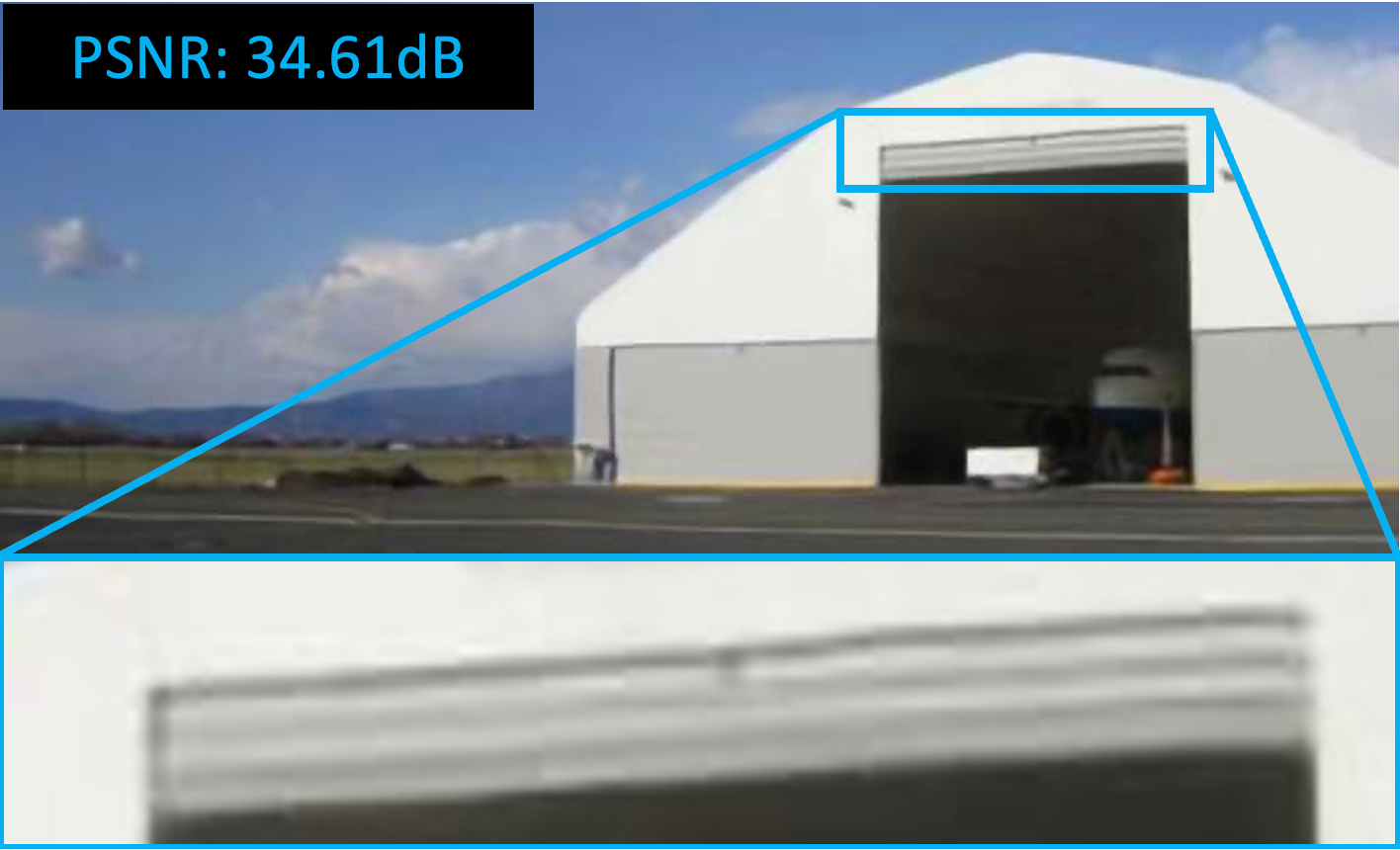}}
  \caption{Example of testing results on our synthetic static scene dataset. (a). Input (b).Ground truth (c). Output of {\bf{TMT [ours]}} (d).The output of TSRWGAN \cite{Jin2021NatureMI} (e). The output of VRT \cite{liang2022vrt} (f). Output of BasicVSR++ \cite{chan2022basicvsrpp}}
  \label{fig:static_example}
\end{figure}

\subsection{Quantitative Comparison With Other Networks}
The neural networks we trained are evaluated on our synthetic testing set. We measured PSNR, SSIM, complex-wavelet SSIM (CW-SSIM) \cite{sampat2009complex} and perceptual quality LPIPS \cite{zhang2018unreasonable} for quantitative comparison. The testing results are shown in Table \ref{table:tab1} and \ref{table:tab2}. The TMT outperforms the multi-frame turbulence mitigation network TSRWGAN \cite{Jin2021NatureMI} and 2 SOTA video restoration networks BasicVSR++ \cite{chan2022basicvsrpp} and VRT \cite{liang2022vrt} by a large margin. We noticed the CW-SSIM scores are closer to perceptual quality, and all metrics are mostly correlated when we do not use perceptual loss. We also provide a visual comparison in Fig. \ref{fig:static_example} to show the advance of our method.

\begin{table}[h]
\centering
\setlength{\aboverulesep}{0.3pt}
\setlength{\belowrulesep}{0.3pt}
\caption{Comparison on the {\bf{STATIC}} scene dataset.}
\label{table:tab1}
\begin{tabular}{lcccc}
\toprule[1pt]
Methods / \# frames & PSNR & SSIM & CW-SSIM & LPIPS($\downarrow$)\\
\midrule
TurbNet \cite{mao2022single} / 1 & 22.7628 & 0.6923 & 0.8230  & 0.4012 \\
\midrule
BasicVSR++ \cite{chan2022generalization} / 12 & 26.5055 & 0.8121 & 0.9189  & 0.2587 \\
TSRWGAN \cite{Jin2021NatureMI} / 15  & 25.2888 & 0.7784 & 0.8982  & 0.2243 \\
VRT \cite{liang2022vrt} / 12   & 27.4556 & 0.8287 & 0.9338 & 0.1877\\
\midrule
TMT  [ours] / 12 & \underline{27.7309} & \underline{0.8341} & \underline{0.9376} & \underline{0.1815} \\
TMT [ours] / 20  & \textbf{28.4421} & \textbf{0.8580} & \textbf{0.9452} & \textbf{0.1693} \\
\bottomrule[1pt]
\end{tabular}
\end{table}

\begin{table}[ht]
\centering
\setlength{\aboverulesep}{0.3pt}
\setlength{\belowrulesep}{0.3pt}
\caption{Comparison on the {\bf{DYNAMIC}} scene dataset.}
\label{table:tab2}
\begin{tabular}{lcccc}
\toprule[1pt]
Methods / \# frames & PSNR & SSIM & CW-SSIM & LPIPS($\downarrow$)\\
\midrule
TurbNet \cite{mao2022single} / 1  & 24.2229 & 0.7149 & 0.8072 & 0.4445 \\
\midrule
BasicVSR++ \cite{chan2022generalization} / 12 & 27.0231 & 0.8073 & 0.8653  & 0.2492\\
TSRWGAN \cite{Jin2021NatureMI} / 15  & 26.3262 & 0.7957 & 0.8596  & 0.2606 \\
VRT \cite{liang2022vrt} / 12   & 27.6114 & 0.8300 & 0.8691  & 0.2485 \\
\midrule
TMT [ours] / 12  & \underline{27.8816} & \underline{0.8318} & \underline{0.8705}  & \underline{0.2475} \\
TMT [ours] / 20  & \textbf{28.0124} & \textbf{0.8352} & \textbf{0.8741} & \textbf{0.2412} \\
\bottomrule[1pt]
\end{tabular}
\end{table}

We present the inference-time computing budget of all models we trained in Table \ref{table:computation_comparison}. The measurement is based on a single NVIDIA 2080 Ti GPU. If the full size cannot fit into the GPU, we split the input into the largest possible patches. The splitting should be overlapped to reduce artifacts. Although VRT has the closest performance in terms of PSNR and SSIM to TMT, TMT requires much less computation and memory requirement; hence, it is much more efficient.
\begin{table}[ht]
  \caption{Comparison of computational consumption in inference. The speed is measured per frame on $540 \times 960$ resolution images. The first two rows are conventional methods tested on CPU.}
  \label{table:computation_comparison}
  \centering
  \setlength{\aboverulesep}{0.2pt}
\setlength{\belowrulesep}{0.2pt}
  \resizebox{0.98\linewidth}{!}{\setlength\doublerulesep{0.5pt}
  \begin{tabular}{lccc}
    \toprule[1pt]
     Methods    & $\#$ parameters (M)  & FLOPs/frame (G) & speed (s) \\
    \midrule
     Mao et al. \cite{mao_tci}    & - & - & $\sim 5500$ \\
     CLEAR \cite{Anantrasirichai2013}   & - & - & $\sim 20$ \\
    \midrule
    BasicVSR++ \cite{chan2022generalization}  & 9.76  & $338.4$  & $0.08$  \\
    TSRWGAN \cite{Jin2021NatureMI}   & 46.28  & $2471$  & $1.15$ \\
    VRT \cite{liang2022vrt}  &  18.32 & $7756$ & $5.88$ \\
    TMT [ours] & 26.04  & $1826$ & $1.52$  \\
    \bottomrule[1pt]
  \end{tabular}}
\end{table}

\subsection{Ablation Study}
\begin{table*}[ht]
  \caption{Quantitative comparison on the synthetic testing set for different design choices of the TMT. $\text{PSNR}_{Y}$ and $\text{SSIM}_{Y}$ indicate PSNR and SSIM measured in the YCbCr space. $\text{TMT}_{a}$ and $\text{TMT}_{b}$ denote the TMT model with variants a and b of the attention module, respectively. ``warp'' means the tilt-removal module, ``SS'' and ``MS'' are aberrations of single-scale input and multi-scale input. FLOPs are measured with an input size of $12 \times 208 \times 208$. All networks are trained with batch size 1 and 800K iterations.}
  \label{table:ablation}
  \centering
  \setlength{\aboverulesep}{0.2pt}
\setlength{\belowrulesep}{0.2pt}
  \begin{tabular}{ccccccccccc}
    \toprule[1pt]
    Dataset & \multicolumn{4}{c}{Static Scenes} & \multicolumn{4}{c}{Dynamic Scenes}  & \multicolumn{2}{c}{Computation Consumption}   \\
    \cmidrule(r){1-1} \cmidrule(r){2-5} \cmidrule(r){6-9} \cmidrule(r){10-11}
    Methods     & PSNR & SSIM & $\text{PSNR}_{Y}$ & $\text{SSIM}_{Y}$  & PSNR & SSIM & $\text{PSNR}_{Y}$ & $\text{SSIM}_{Y}$ & $\#$ Params (M) & FLOPs (G)\\
    \midrule
    $\text{SS-TMT}_{a}$ w.o. warp  & 27.2782 & 0.8221 & 28.7316  & 0.8398 & 27.5635 & 0.8258 & 29.0644  & 0.8459  & \textbf{22.75}  & 1490 \\
    $\text{SS-TMT}_{b}$ w.o. warp  & 27.3092 & 0.8235 & 28.7626  & 0.8411 & 27.5864 & 0.8265 & 29.0886  & 0.8465 & 22.95  & 1514 \\
    $\text{SS-TMT}_{a}$  & 27.3432 & 0.8257 & 28.8013 & 0.8431 & 27.5779 & 0.8282 & 29.0863  & 0.8487 & 24.87  & 1676 \\
    $\text{SS-TMT}_{b}$  & 27.3836 & 0.8266 & 28.8396 & 0.8440 & 27.6051 & 0.8281 & 29.1110  & 0.8485 & 25.07  & 1700 \\
    $\text{MS-TMT}_{a}$ w.o. warp  & 27.4718 & 0.8291 & 28.9243  & 0.8452 & 27.6637 & 0.8301  & 29.1685  & 0.8501  & 23.70  & \textbf{1206} \\
    $\text{MS-TMT}_{b}$ w.o. warp  & 27.5003 & 0.8301 & 28.9602  & 0.8468 & 27.6841 & 0.8310 & 29.1894  & 0.8506 & 23.92 & 1304 \\
    $\text{MS-TMT}_{a}$  & 27.5215 & 0.8307 & 28.9781 & 0.8469 & 27.7239  & \textbf{0.8329}  & 29.2337   & \textbf{0.8526}  & 25.82  & 1392 \\
    \midrule[0.3pt]
    $\text{MS-TMT}_{b}$ [Proposed]  & \textbf{27.5422} & \textbf{0.8320} & \textbf{29.0011} & \textbf{0.8487} & \textbf{27.7419} & 0.8323 & \textbf{29.2510}  & 0.8520 & 26.04  & 1490\\
    \bottomrule[1pt]
  \end{tabular}
\end{table*}

\begin{figure}[ht]
    \centering
    \includegraphics[width=0.6\linewidth]{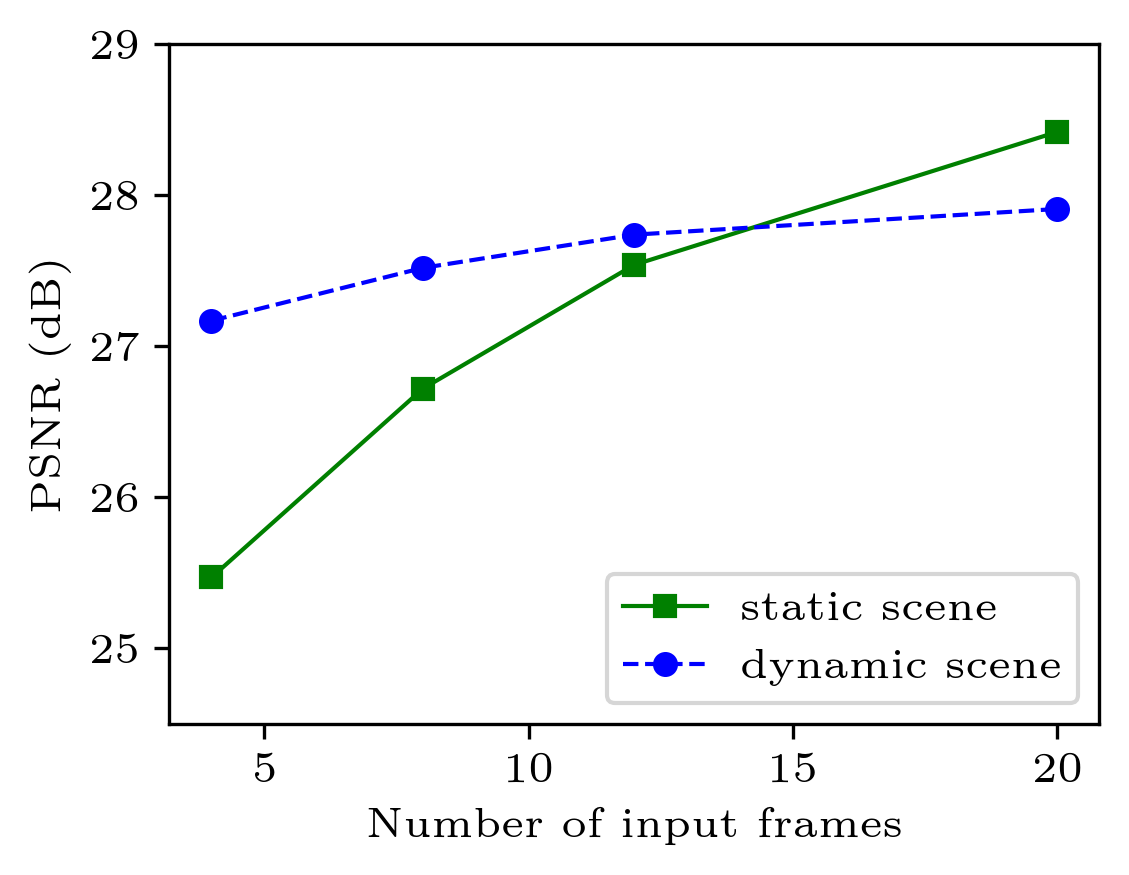}
    \caption{Number of frames and performance}
    \label{fig:nframe}
\end{figure}

\paragraph{Influence of number of input frames}
To restore the clean image sequence under turbulence, the network must perceive a large enough area in the spatial-temporal domain to estimate the property of the degradation field, especially for samples having high temporal correlation. If the correlation factor is 1, all frames have the same degradation, the multi-frame reconstruction will collapse into a single-frame problem, which is extremely hard to solve, as revealed by \cite{vint2020analysis}. To understand how the number of frames would affect the quality of restoration, we trained our $\text{TMT}_{b}$ using 4, 8, 12, and 20 input frames with our synthetic dataset. The testing performance is shown in Fig. \ref{fig:nframe}. The results show the performance of the network can be improved with more input frames. We also note that to make a fair comparison with a strong baseline such as \cite{liang2022vrt}, which requires large GPU memory space so that it is hard to deploy with more than 12 frames, we kept a 12-frame setting in this paper. However, we can easily train with 20 frames to get a much better performance than VRT with the same hardware.

\paragraph{Two-stage design and the tilt-removal module}
To evaluate the value of the tilt-removal module, we trained our TMT backbone to restore images directly without removing the tilt. The result is shown in Table \ref{table:ablation}. As it shows, the tilt-removal module can boost the overall performance by $0.02 \sim 0.08$dB. Tilt-removal is more effective in static scenes, as distinguishing true object motion from turbulence distortions in dynamic scenes introduces uncertainty. Note the tilt-removal module is very lightweight. If we scale up one-stage TMT, we can only add 10\% more channels on the one-stage TMT and finally get around 0.01 dB improvement, much less than the two-stage model. Moreover, instead of more parameters, more frames have a more significant impact on the performance. The two-stage model can be scaled up to take more frames because they are applied sequentially. The one-stage model has a lower frame number limitation than the two-stage model, further restricting the performance.

\paragraph{Influence of channel shuffle operation}
The purpose of the channel shuffle operation is to facilitate the communication of channels across different frames in the TCJA module. The comparison in Table \ref{table:ablation} demonstrates the effectiveness of the channel shuffle operation. The performance benefits $0.02 \sim 0.03$ dB from the shuffle operation compared to the plain connection.

\paragraph{Influence of multi-scale input in the TMT backbone}
For general U-shaped design \cite{ronneberger2015u}, only a single input is required to feed into the encoder. Several works \cite{Nah2017cvpr, ji2022multi, zamir2021multi} suggest feeding inputs in multiple scales/resolutions could help the network to learn more efficiently. In the proposed paper, we utilize multi-scale input by simply downsampling the raw input into lower resolutions. To demonstrate the advantage of this operation, we also trained our TMT with single-scale input: the input dimension of each level in the encoder remains the same, only the input feature is changed from the concatenation of outputs produced by the last level, and the 3D convolution layer to solely from the last level. The comparison in Table \ref{table:ablation} demonstrates the effectiveness of the multi-scale input. The PSNR of networks with multi-scale input is $\sim 0.15$ dB higher than that with single-scale input, while the consumption complexity does not increase.

\subsection{Generalization To Real-World Data}
With our physically-based simulation process, samples of real-world turbulence can be viewed as the interpolation of samples generated by our simulation. Other simulation methods are either much more time-consuming or inaccurate \cite{mao_P2S}. Two recent data-driven works, TSRWGAN \cite{Jin2021NatureMI} and complex CNN \cite{anantrasirichai2022atmospheric}, also used synthetic data to train neural networks, but their generalization capability is limited. Quantitative comparisons among models on real-world images are hard because most real-world image sequences don't have ground truth. Despite this problem, direct visual comparison can still illustrate how our synthetic data helps generalization.

\begin{figure*}[ht]
    \captionsetup[subfloat]{farskip=2pt, font=scriptsize}
    \centering
    \includegraphics[width=0.42\linewidth]{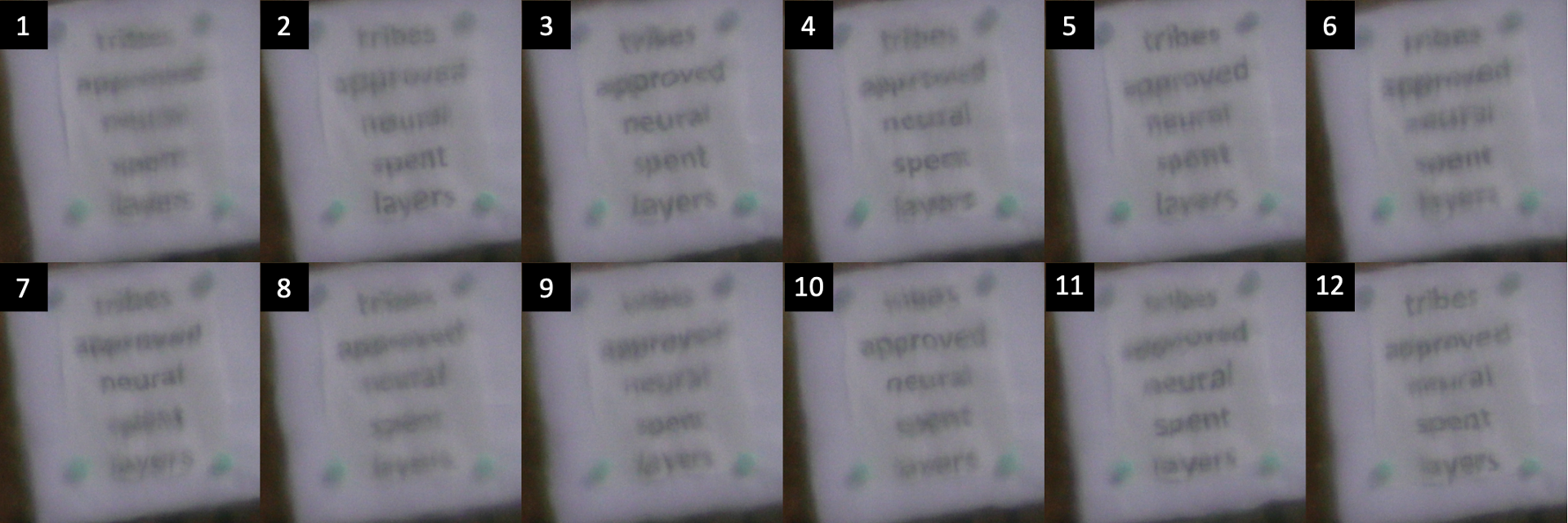}%
    \hfill
    \includegraphics[width=0.14\linewidth]{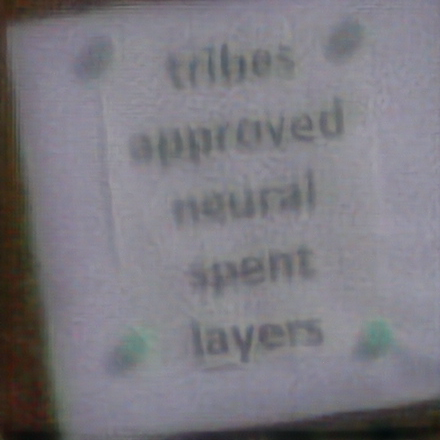}%
    \hfill
    \includegraphics[width=0.14\linewidth]{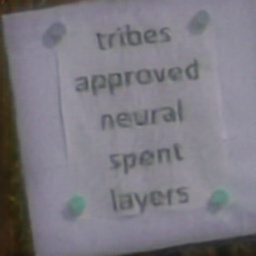}%
    \hfill
    \includegraphics[width=0.14\linewidth]{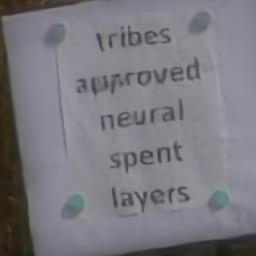}%
    \hfill
    \includegraphics[width=0.14\linewidth]{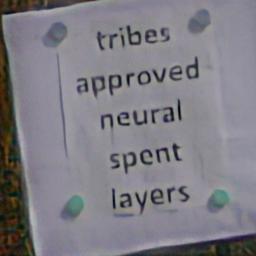}%
    \\
    \vspace{2pt}
    \includegraphics[width=0.42\linewidth]{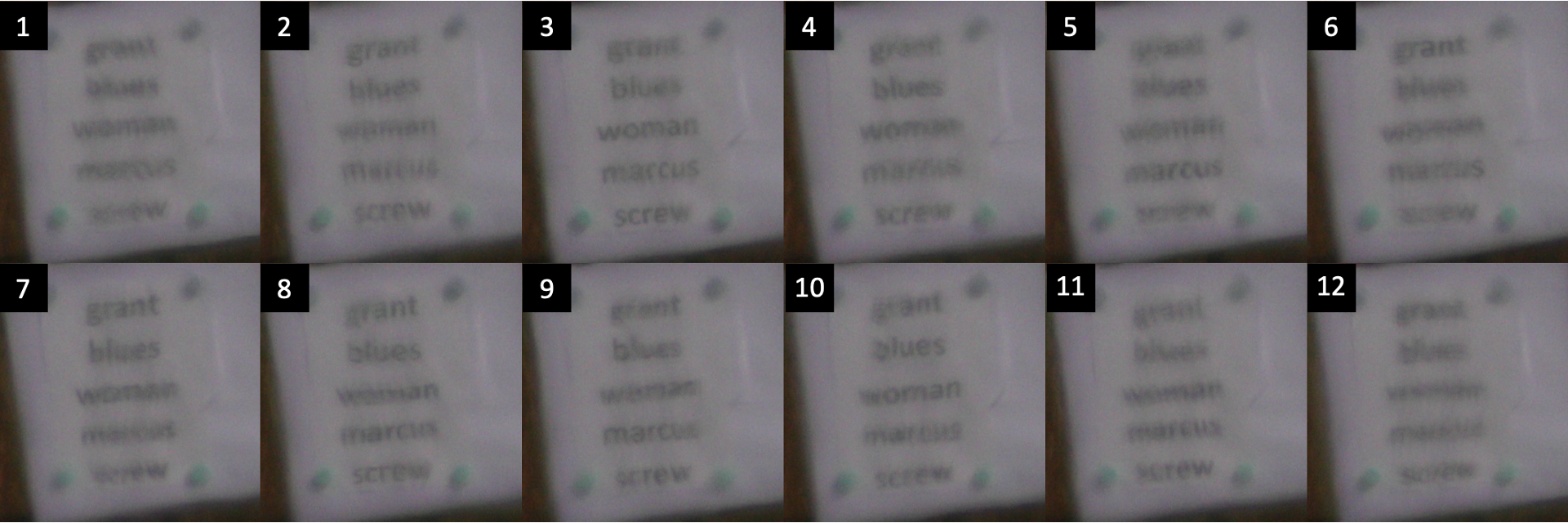}%
    \hfill
    \includegraphics[width=0.14\linewidth]{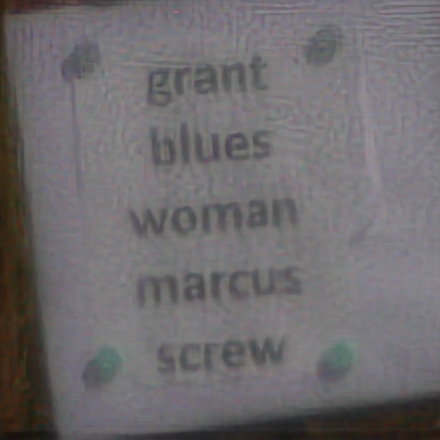}%
    \hfill
    \includegraphics[width=0.14\linewidth]{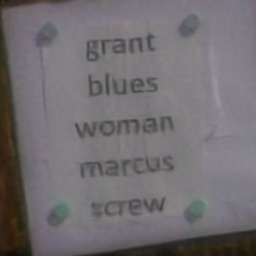}%
    \hfill
    \includegraphics[width=0.14\linewidth]{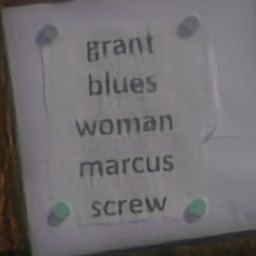}%
    \hfill
    \includegraphics[width=0.14\linewidth]{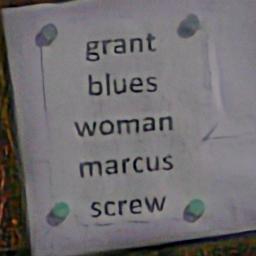}%
    \\
  \subfloat[Input sequences]{%
      \includegraphics[width=0.42\linewidth]{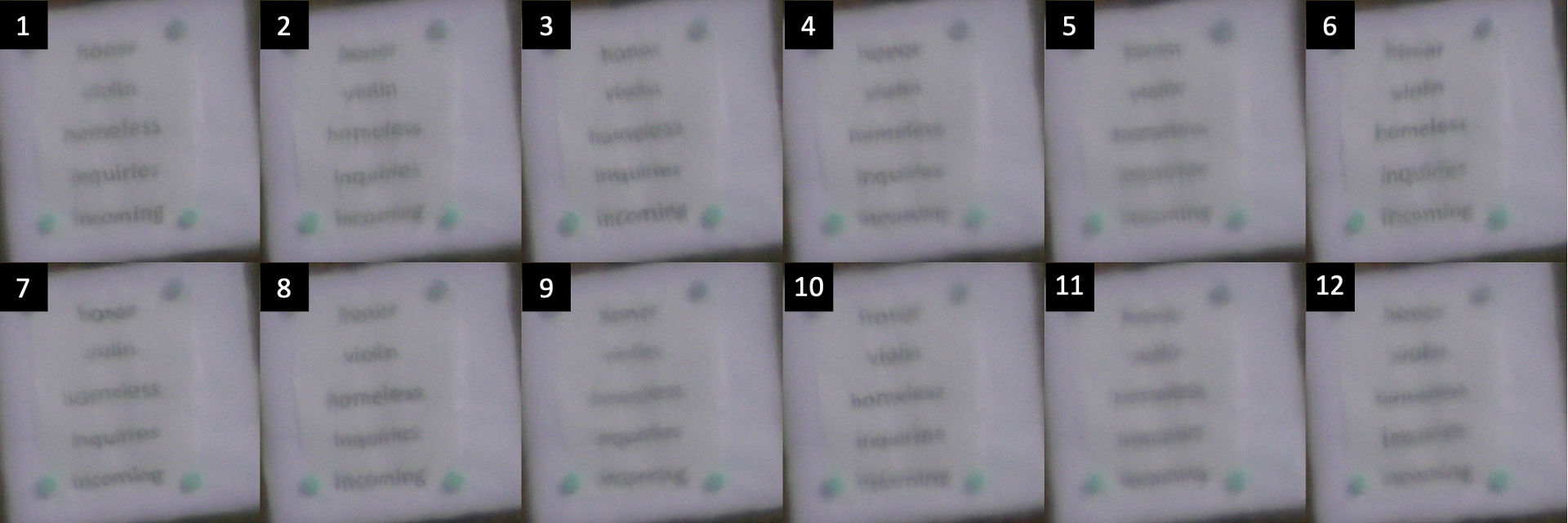}}
    \hfill
  \subfloat[BasicVSR++ \cite{chan2022basicvsrpp}]{%
    \includegraphics[width=0.14\linewidth]{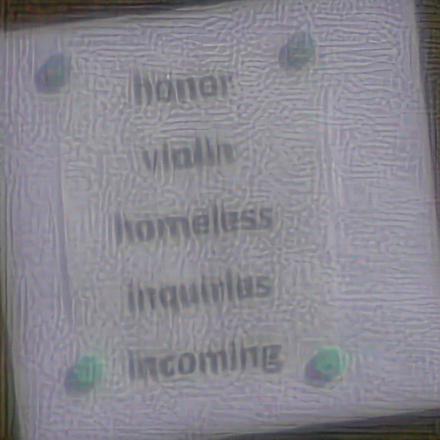}}
   \hfill
  \subfloat[TSRWGAN \cite{Jin2021NatureMI}]{%
        \includegraphics[width=0.14\linewidth]{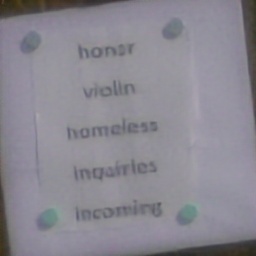}}
    \hfill
  \subfloat[VRT \cite{liang2022vrt}]{%
        \includegraphics[width=0.14\linewidth]{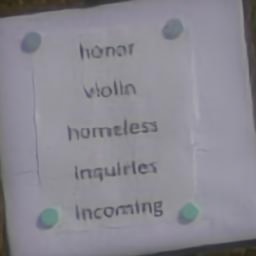}}
    \hfill
  \subfloat[{TMT [ours]}]{%
        \includegraphics[width=0.14\linewidth]{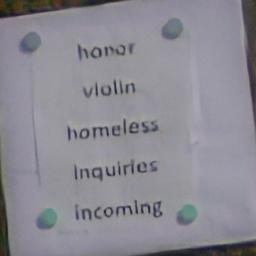}}
  \caption{\textbf{Test on sequences of real-world text data \cite{UG2}}. The three lines of images from top to bottom are the input and output of the 2nd, 24th, and 96th sequences in that dataset. Column (a). Input sequences (b). The output of BasicVSR++ \cite{chan2022basicvsrpp} (c). The output of TSRWGAN \cite{Jin2021NatureMI} (d).VRT \cite{liang2022vrt} (e).TMT [ours]}
  \label{fig:text}
\end{figure*}

\paragraph{Visual comparison among models trained on our datasets}
In the CVPR 2022 UG2+ Challenge, a new long-range turbulence dataset is released for benchmarking turbulence mitigation algorithms \cite{UG2}. This dataset consists of 100 image sequences of text patterns captured from 300 meters away in hot weather. We tested our trained models on this dataset. All models have a certain generalization capability, but our network performs better than others under heavy turbulence. Several samples are given in Fig. \ref{fig:text}.

\begin{figure*}[htbp]
    \captionsetup[subfloat]{farskip=2pt, font=scriptsize}
    \centering
  \subfloat[Input]{%
    \includegraphics[width=0.33\linewidth, height=0.165\linewidth]{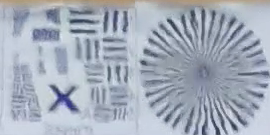}}
    \hfill
  \subfloat[Ground truth]{%
    \includegraphics[width=0.33\linewidth, height=0.165\linewidth]{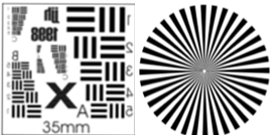}}
   \hfill
  \subfloat[BasicVSR++ \cite{chan2022basicvsrpp}]{%
    \includegraphics[width=0.33\linewidth, height=0.165\linewidth]{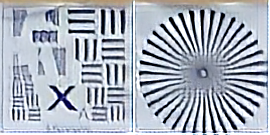}}
   \\
  \subfloat[TSRWGAN \cite{Jin2021NatureMI}]{%
    \includegraphics[width=0.33\linewidth, height=0.165\linewidth]{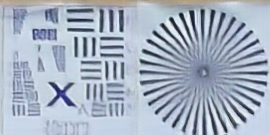}}
   \hfill
  \subfloat[VRT \cite{liang2022vrt}]{%
    \includegraphics[width=0.33\linewidth, height=0.165\linewidth]{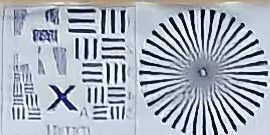}}
   \hfill
  \subfloat[{TMT [ours]}]{%
    \includegraphics[width=0.33\linewidth, height=0.165\linewidth]{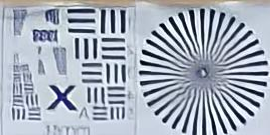}}
  \caption{Visual comparison on the OTIS dataset \cite{gilles2017open}. We show the result of Patterns 15 and 16 which contain the strongest turbulence effect.}
  \label{fig:OTIS}
\end{figure*}

\begin{figure*}[!htbp]
    \centering
    \captionsetup[subfloat]{farskip=2pt, font=scriptsize}
  \subfloat[Input]{%
    \includegraphics[width=0.33\linewidth, height=0.14\linewidth]{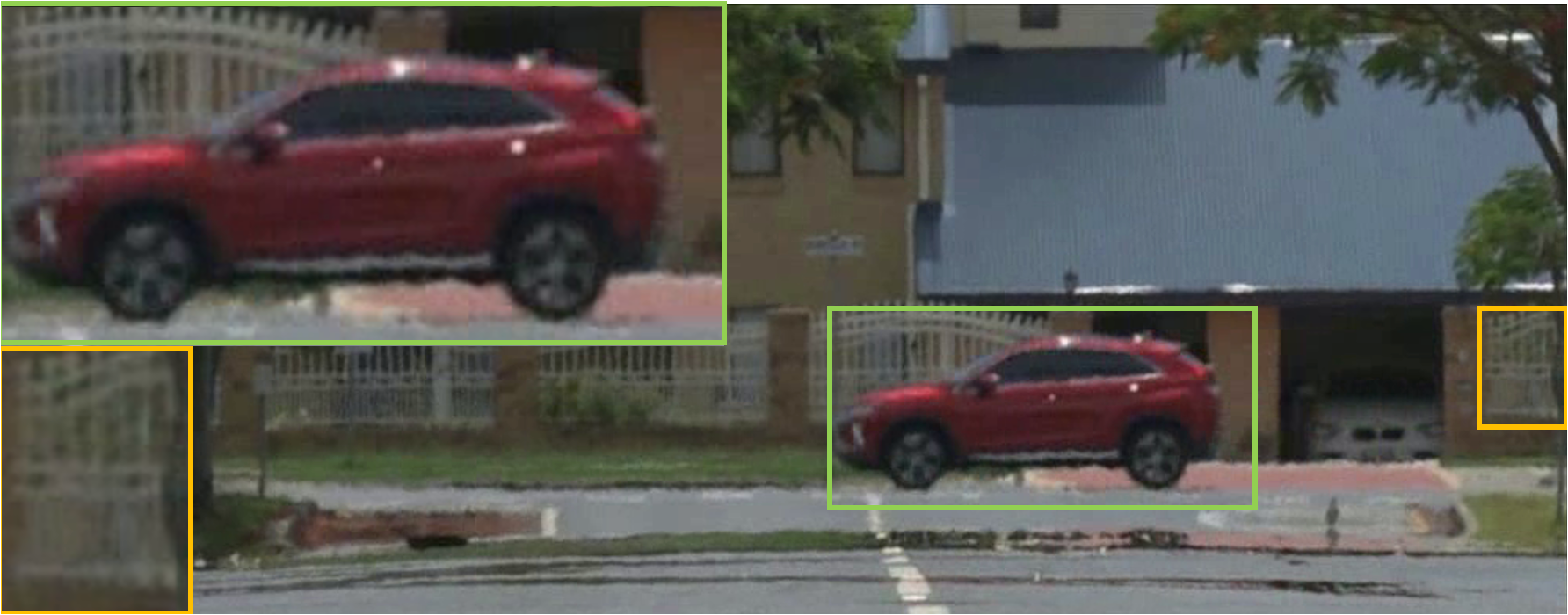}}
    \hfill
  \subfloat[Original WGAN \cite{Jin2021NatureMI}]{%
    \includegraphics[width=0.33\linewidth, height=0.14\linewidth]{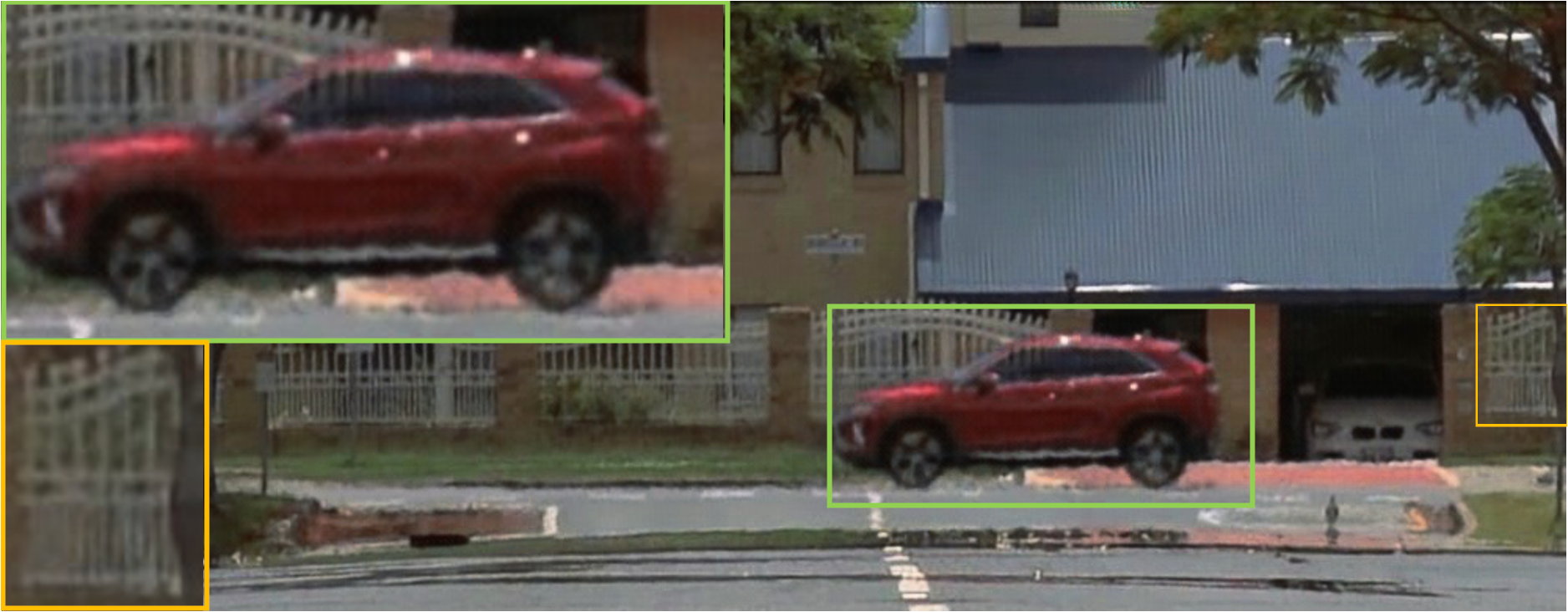}}
   \hfill
  \subfloat[Fine-tuned WGAN]{%
    \includegraphics[width=0.33\linewidth, height=0.14\linewidth]{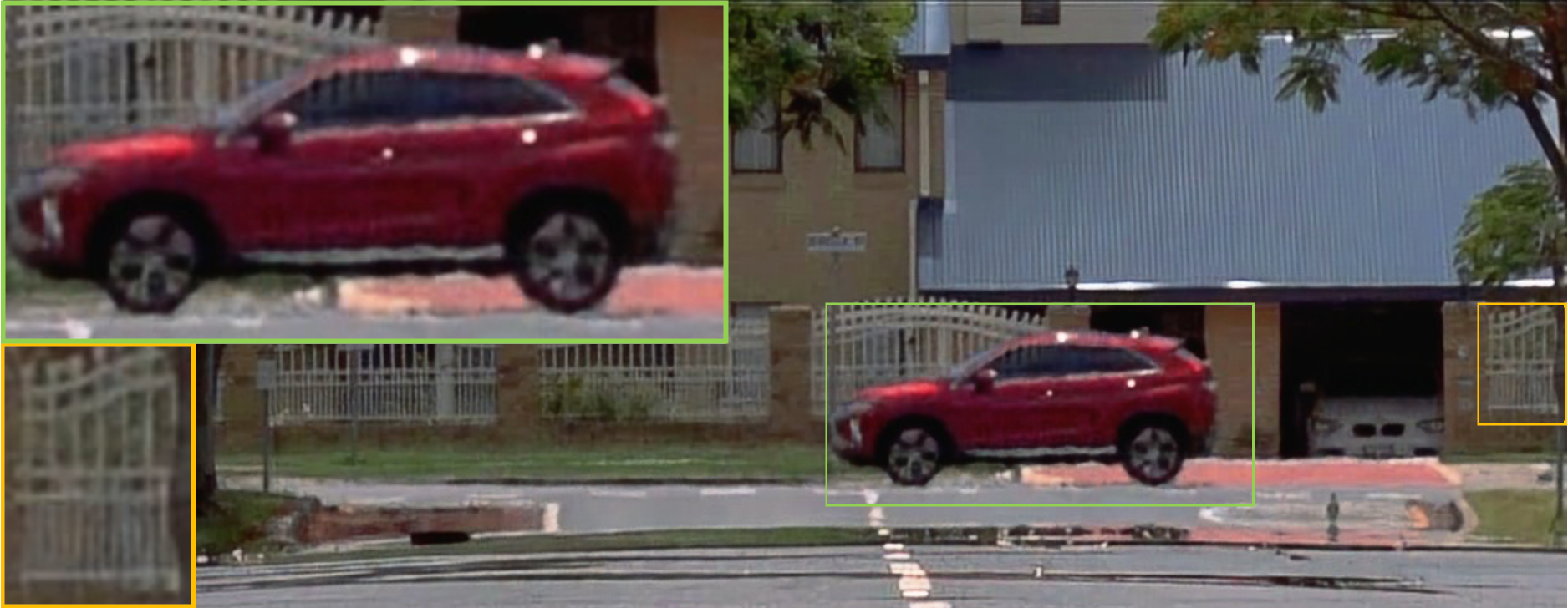}}
    \\
  \subfloat[BasicVSR++ \cite{chan2022basicvsrpp}]{%
    \includegraphics[width=0.33\linewidth, height=0.14\linewidth]{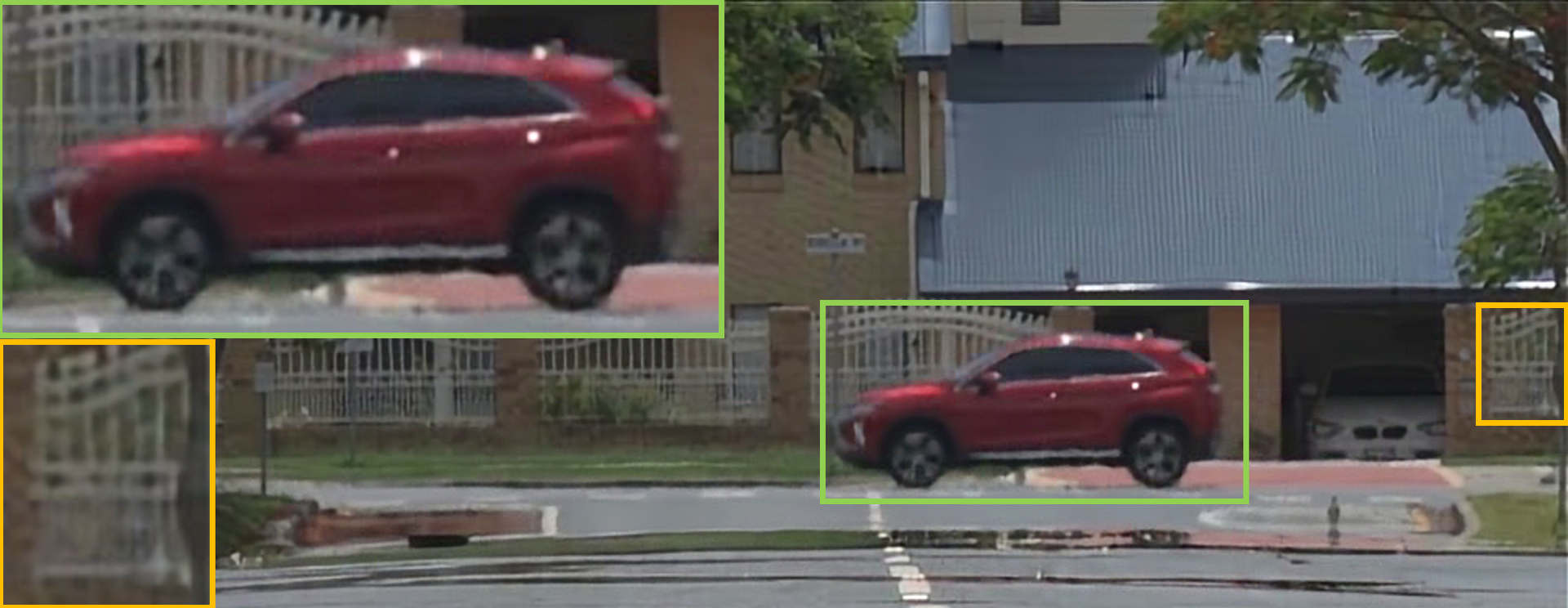}}
    \hfill
  \subfloat[VRT \cite{liang2022vrt}]{%
    \includegraphics[width=0.33\linewidth, height=0.14\linewidth]{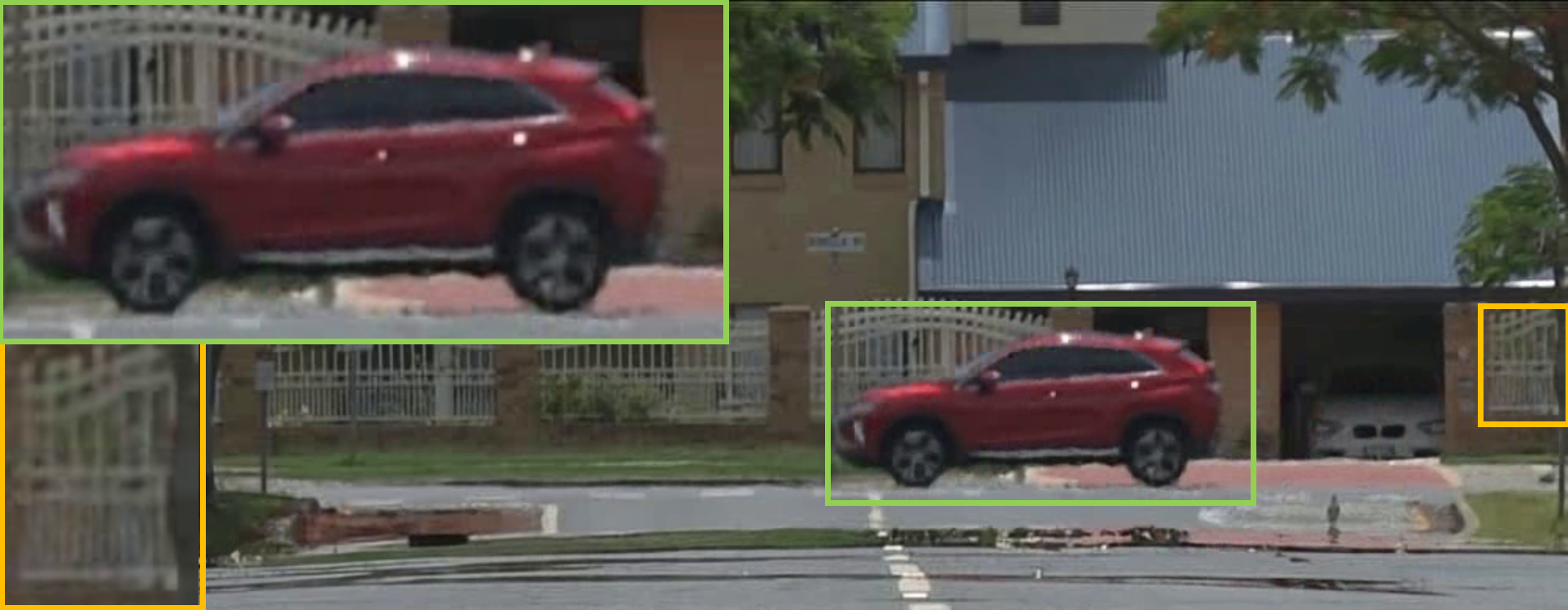}}
   \hfill
  \subfloat[{TMT [ours]}]{%
    \includegraphics[width=0.33\linewidth, height=0.14\linewidth]{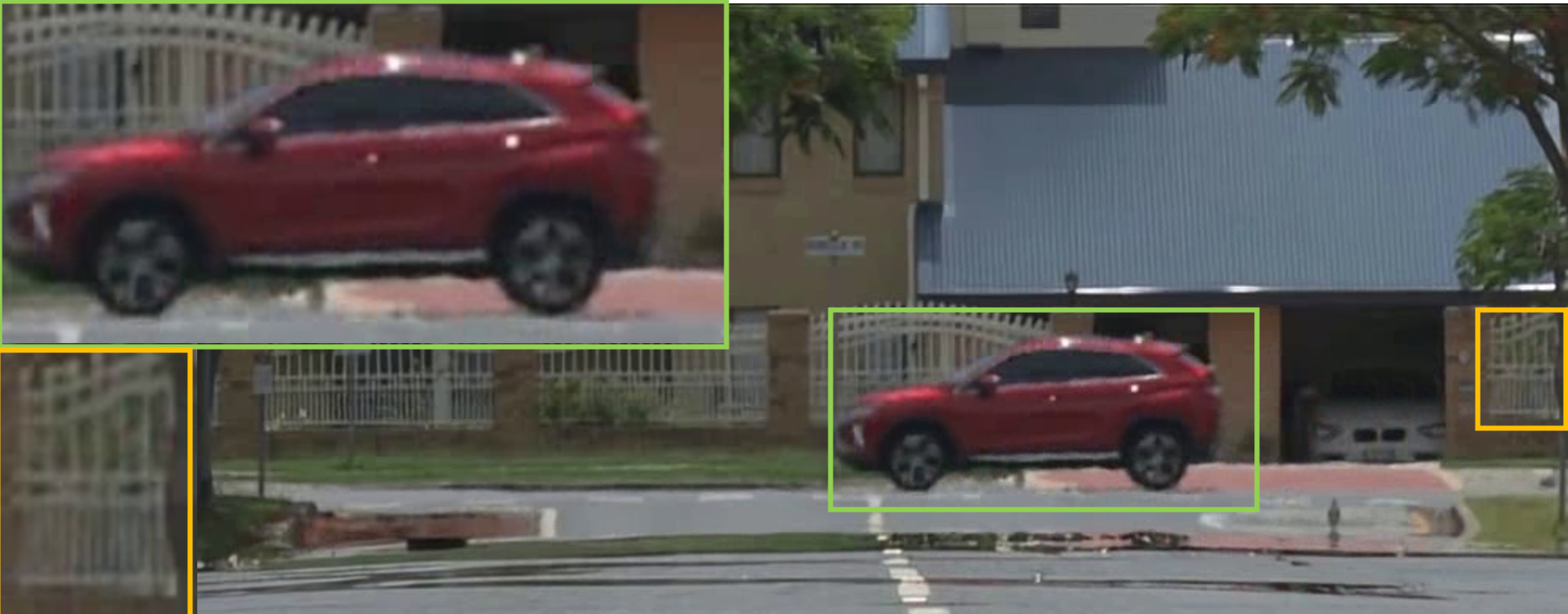}}
  \caption{Visual comparison on real turbulence data from TSRWGAN's dataset \cite{Jin2021NatureMI}. Model of (b) is provided by authors of \cite{Jin2021NatureMI}, and the model for (c) is fine-tuned on our data. From the comparison between (b) and (c), we find the robustness to relatively strong turbulence has been enhanced using our data, despite the artifacts of railing on the window still existing. Notice (f) has the least turbulence left while preserving the most details.}
  \label{fig:TSRWGAN_Data}
\end{figure*}

Besides the turbulence text dataset, we tested all models on an earlier OTIS dataset \cite{gilles2017open} and TSRWGAN's real-world test set. The OTIS has 16 sequences of static patterns in different scales and turbulence levels. The TSRWGAN's real-world test set has 27 dynamic scenes in relatively mild turbulence strength. Since the ground truth is unavailable, we only show some visual comparisons in Fig. \ref{fig:OTIS} and Fig. \ref{fig:TSRWGAN_Data}, from which one can easily conclude that models trained on our dataset generalize well on a broad range of real-world turbulence conditions and among them, our model could restore images in better visual quality than others.

\begin{figure*}[htbp]
    \captionsetup[subfloat]{farskip=2pt, font=scriptsize}
    \centering
  \subfloat[Input]{%
    \includegraphics[width=0.45\linewidth]{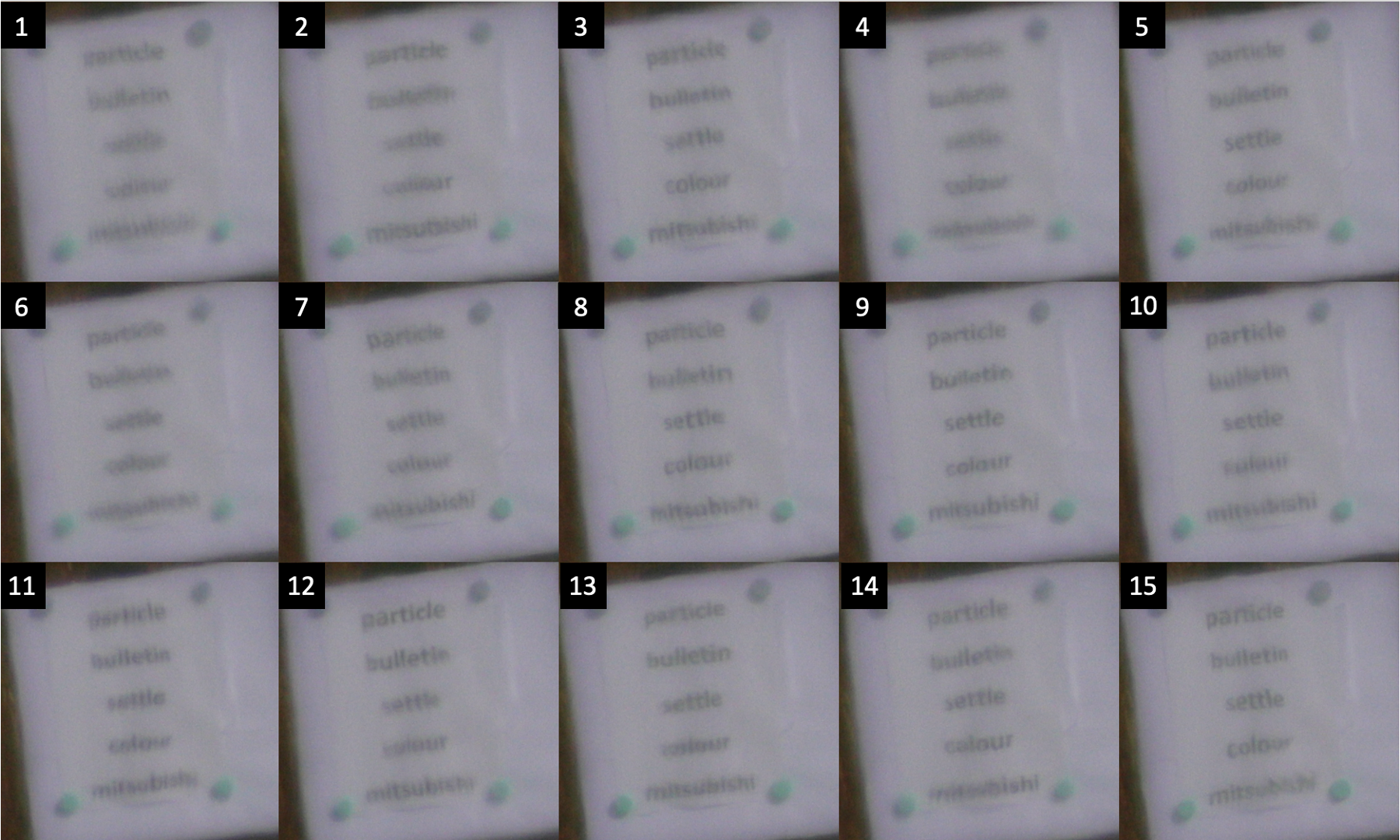}}
    \hfill
  \subfloat[Trained with TSRWGAN's data]{%
    \includegraphics[width=0.27\linewidth]{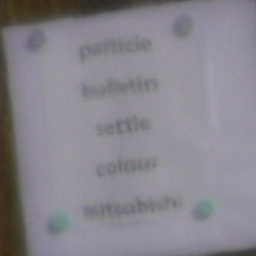}}
   \hfill
  \subfloat[Trained with our data]{%
    \includegraphics[width=0.27\linewidth]{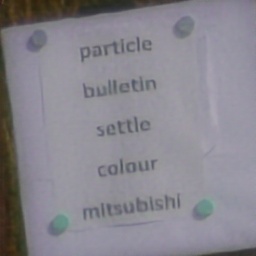}}
    \hfill
    \\
  \subfloat[Input]{%
      \includegraphics[width=0.33\linewidth, height=0.35\linewidth]{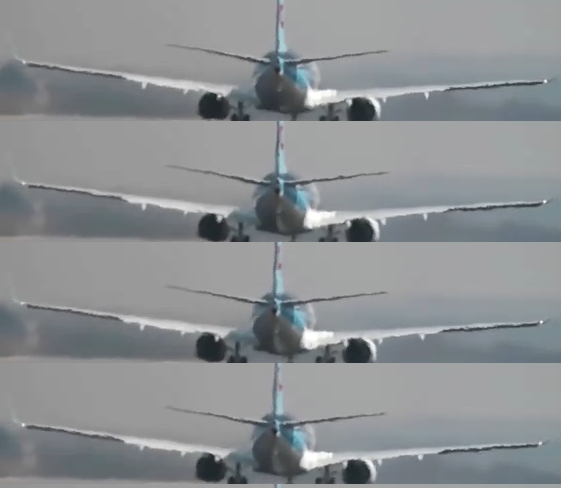}}
    \hfill
  \subfloat[Trained with TSRWGAN's data]{%
    \includegraphics[width=0.33\linewidth, height=0.35\linewidth]{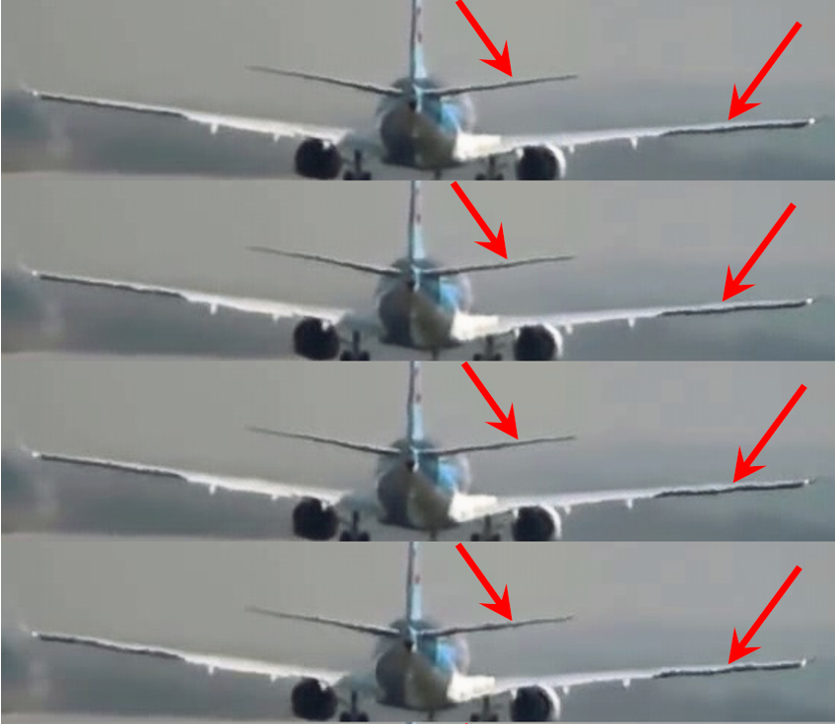}}
   \hfill
  \subfloat[Trained with our data]{%
    \includegraphics[width=0.33\linewidth, height=0.35\linewidth]{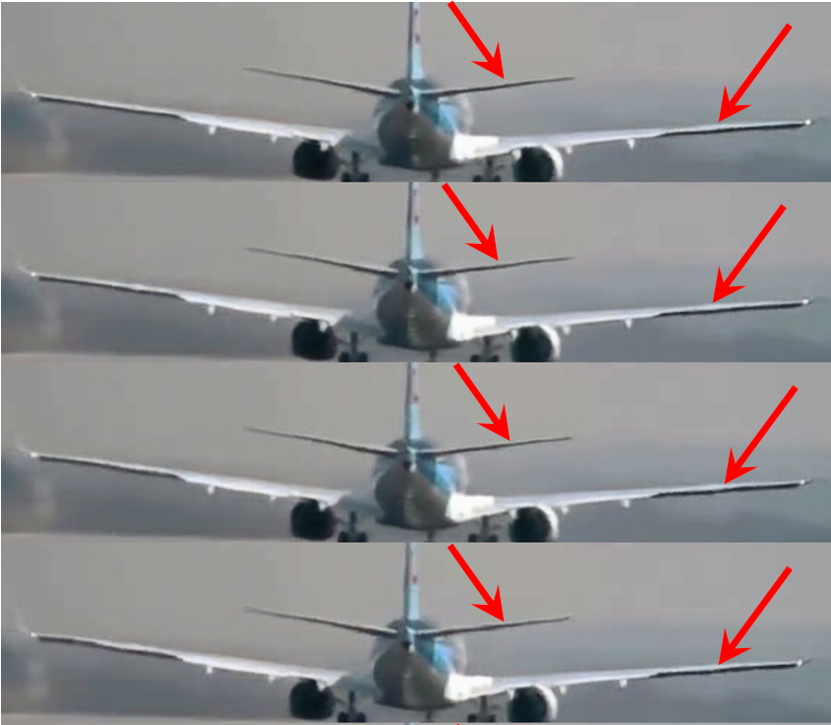}}
    \hfill
  \caption{Test TSRWAGN’s generator on real-world images. (a-c). On text dataset \cite{UG2} (d-f). On CLEAR's dataset (a). Input sequence (d). Single frame sample of input (b, e). Restoration result of TSRWGAN trained with their original data (c, f). Restoration result of TSRWGAN trained with our synthetic datasets. Clearly, our datasets directly facilitate the generalization of real-world images}
  \label{fig:WGAN_comparison}
\end{figure*}

\paragraph{Impact of a better simulator}
The simulation tool that TSRWGAN \cite{Jin2021NatureMI} used is \cite{repasi2011computer}, which generates physically valid tilts, but the blur is spatially invariant. \cite{Jin2021NatureMI} also produces synthetic data by artificial heat sources to create turbulence effects in a short distance. However, this approach tends to generate highly correlated degradation with a weak blur. We observed that their released model does not generalize well on CLEAR's real-world dataset, OTIS \cite{gilles2017open} and the text dataset \cite{UG2} which contain stronger turbulence effect captured at longer range. We fine-tuned TSRWAGN on our synthetic data, showing a significant improvement of the TSRWGAN model on those out-of-distribution datasets in Fig. \ref{fig:WGAN_comparison}.

\begin{figure*}[!htbp]
    \captionsetup[subfloat]{farskip=1pt, font=scriptsize}
    \centering
  \subfloat[Input: $\#12$ frame]{%
    \includegraphics[width=0.248\linewidth]{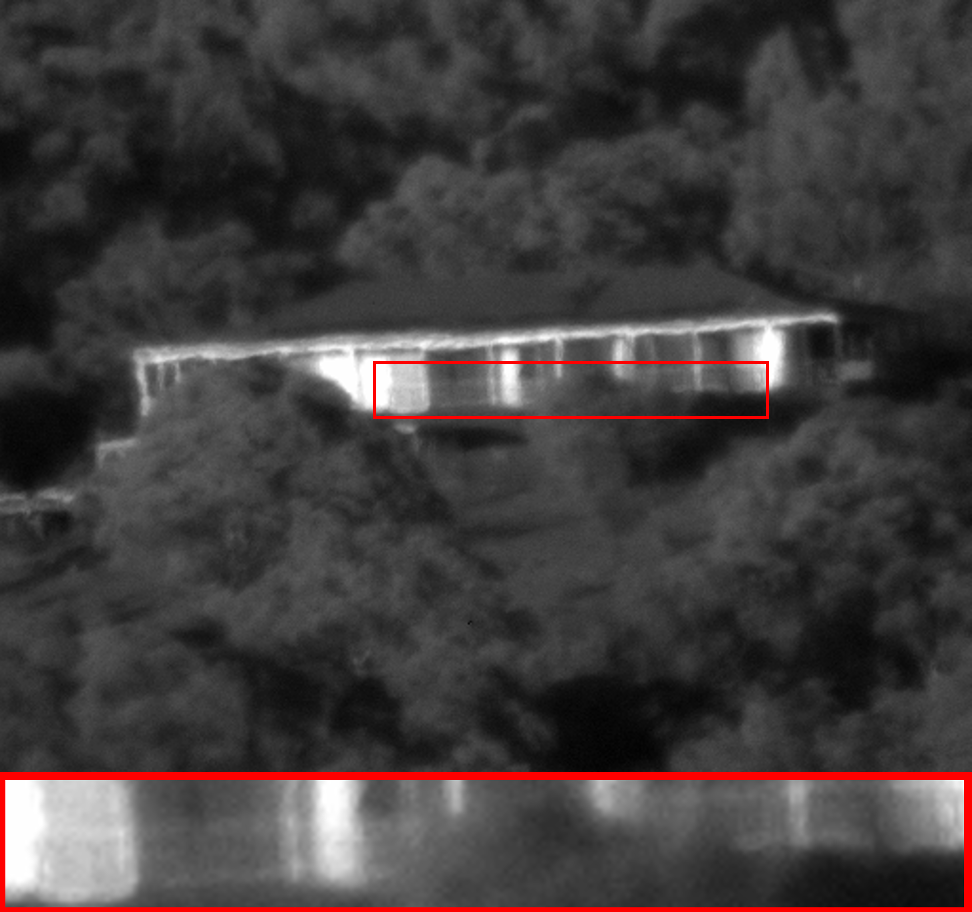}}
    \hfill
  \subfloat[Conventional CLEAR \cite{Anantrasirichai2013}]{%
    \includegraphics[width=0.248\linewidth]{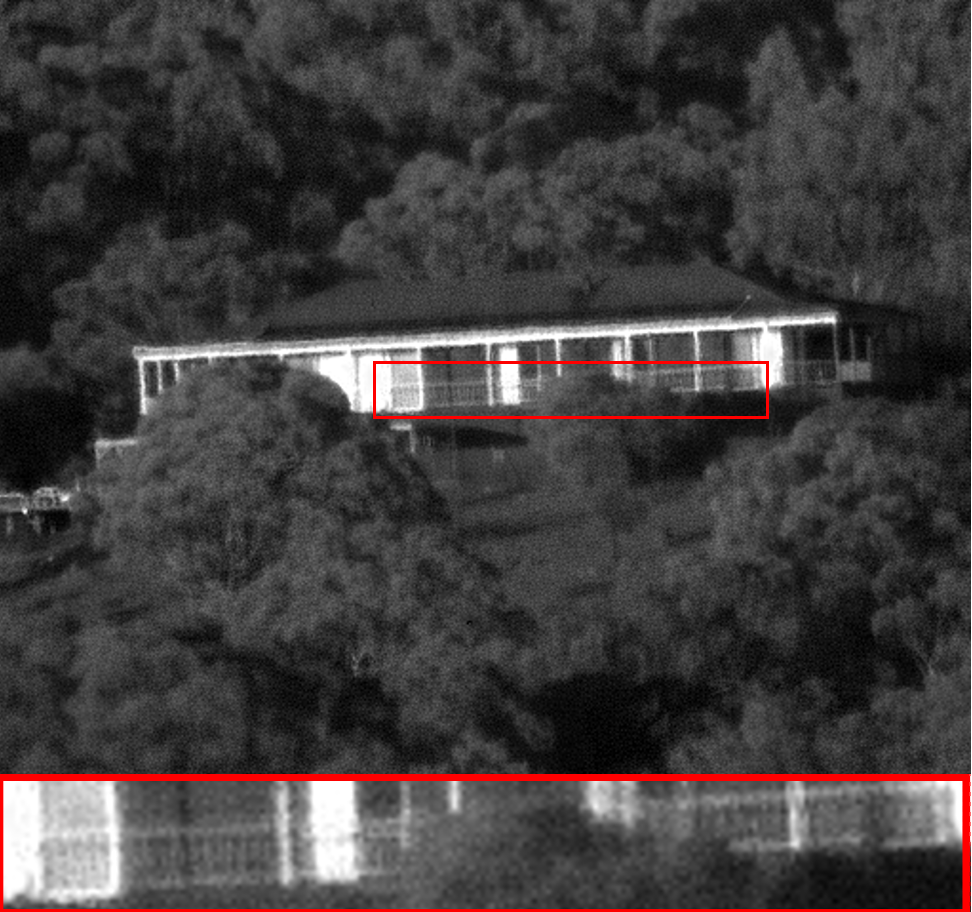}}
   \hfill
  \subfloat[Mao et al. \cite{mao_tci}]{%
    \includegraphics[width=0.248\linewidth]{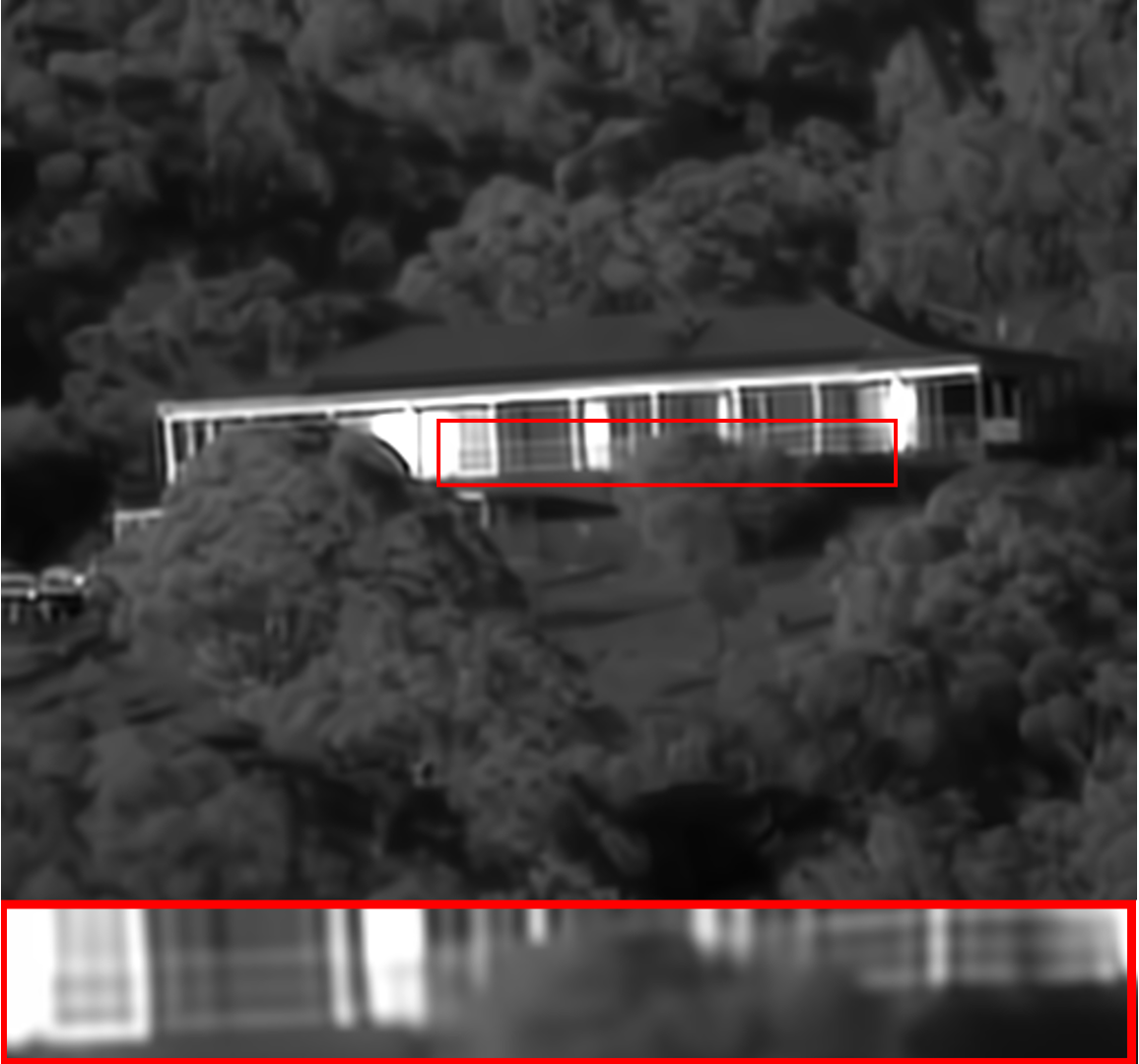}}
   \hfill
  \subfloat[{TMT [ours]}]{%
    \includegraphics[width=0.248\linewidth]{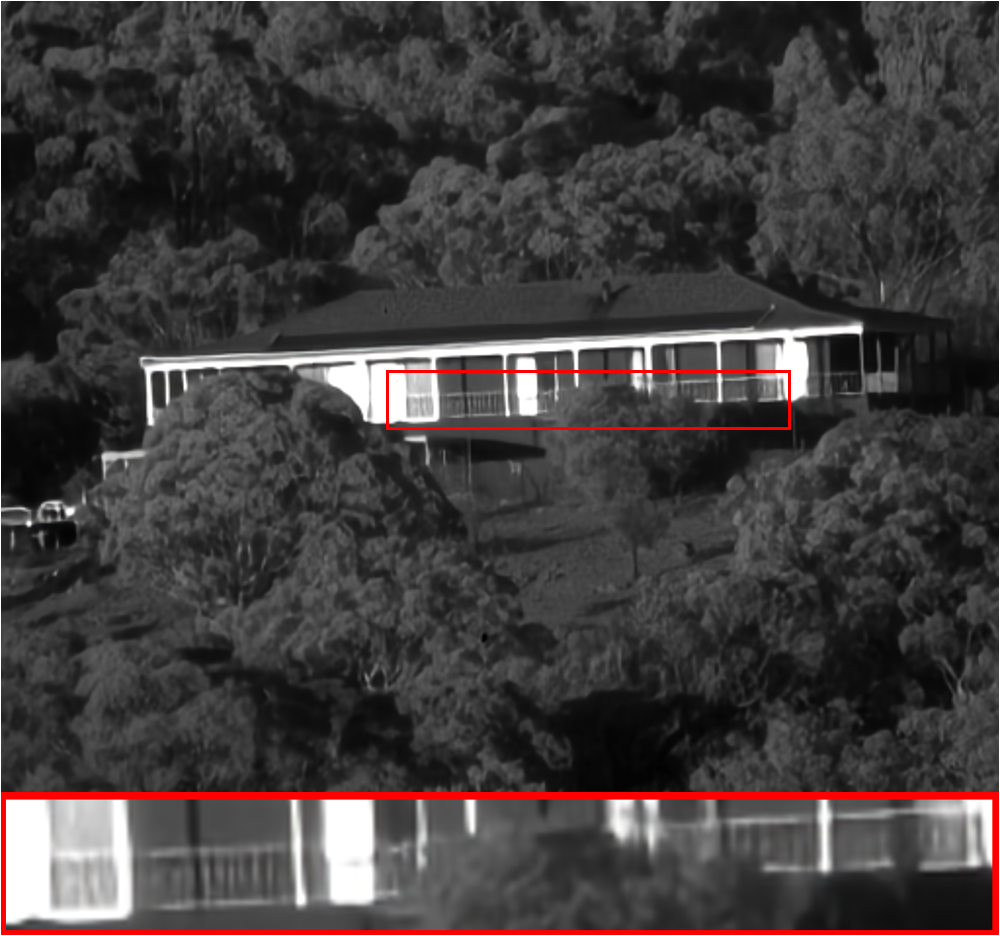}}
  \caption{Comparison with representative conventional methods on the real-world image sequence. Our model is trained on the synthetic static scene dataset. The presented input frame is the 12th in the whole sequence. {\bf{Zoom in for a better view of details}}.}
  \label{fig:Conventional_clear}
\end{figure*}

\begin{figure*}[htbp]
    \captionsetup[subfloat]{farskip=2pt, font=scriptsize}
    \centering
    \includegraphics[width=0.33\linewidth]{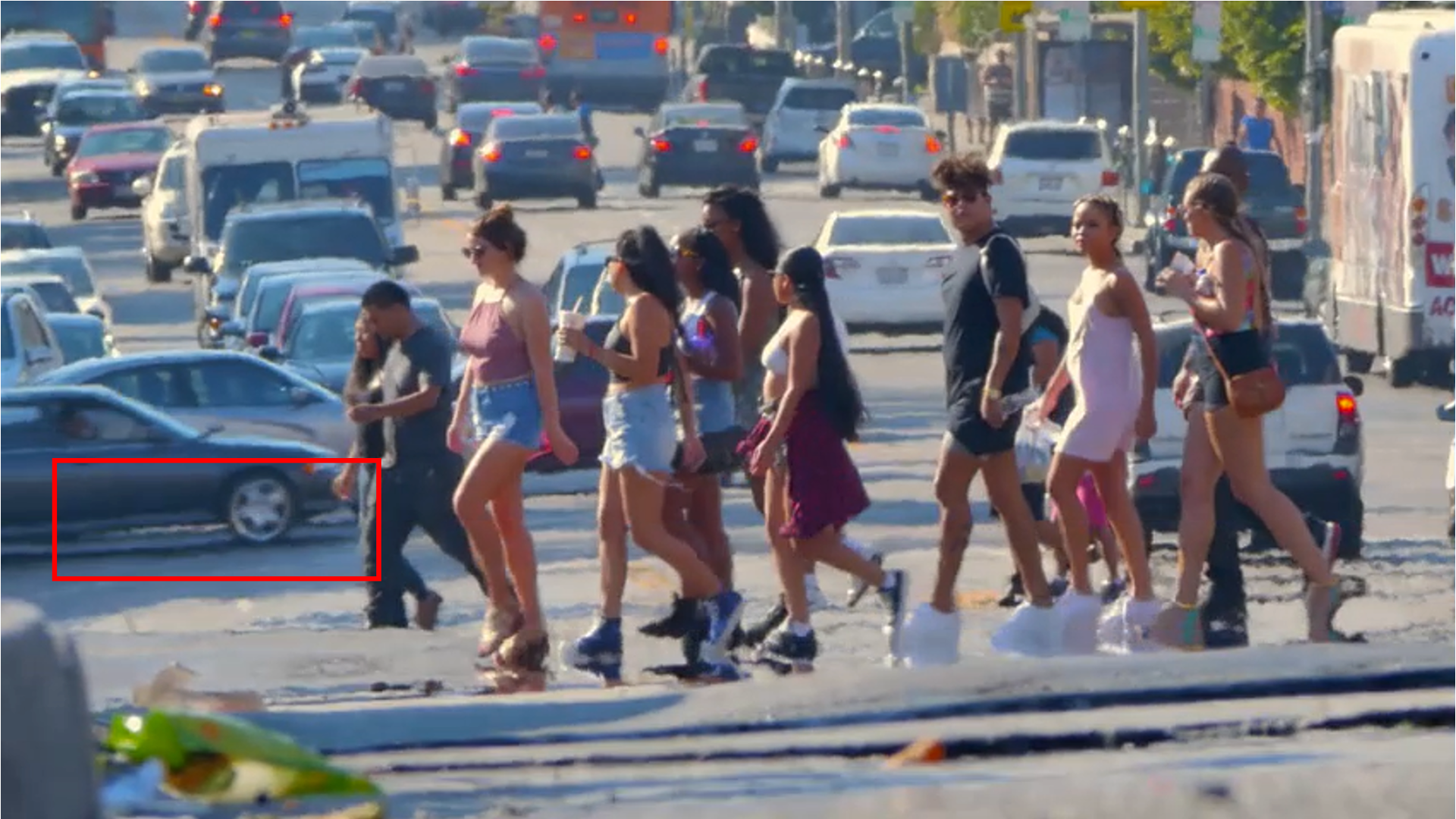}%
    \hfill
    \includegraphics[width=0.33\linewidth]{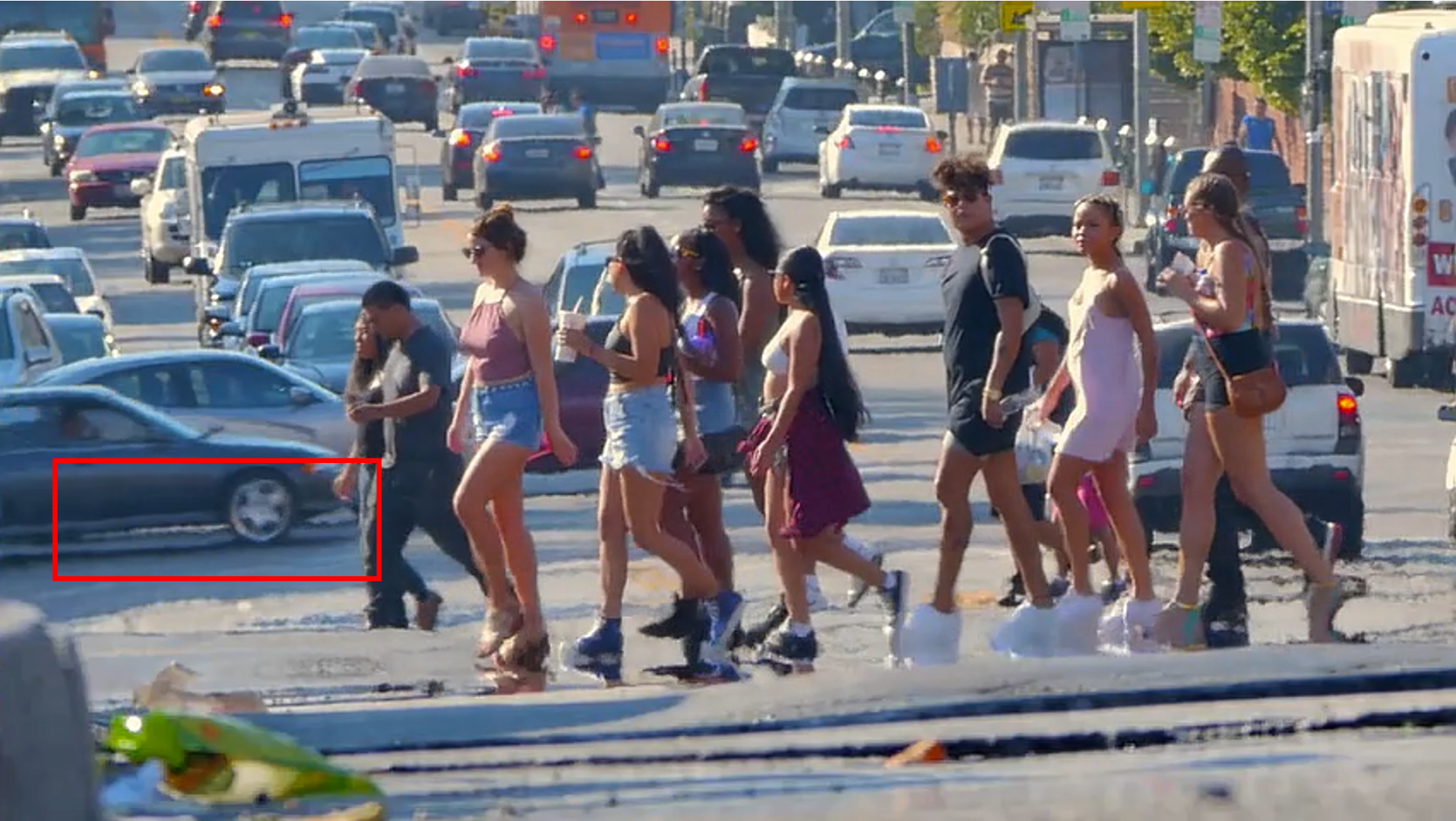}%
    \hfill
    \includegraphics[width=0.33\linewidth]{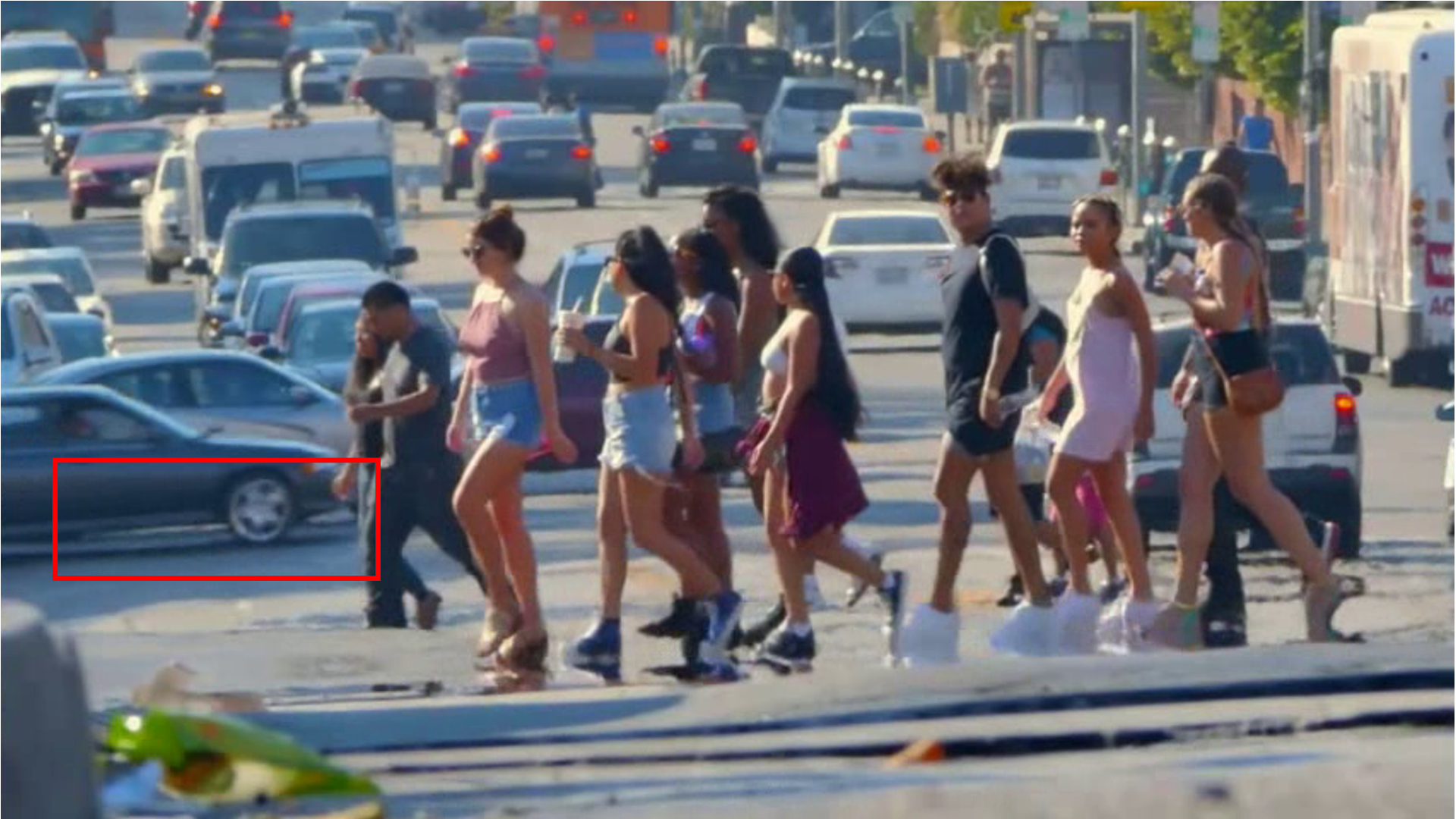}%
    \\
  \subfloat[Input]{%
      \includegraphics[width=0.33\linewidth, height=0.3465\linewidth]{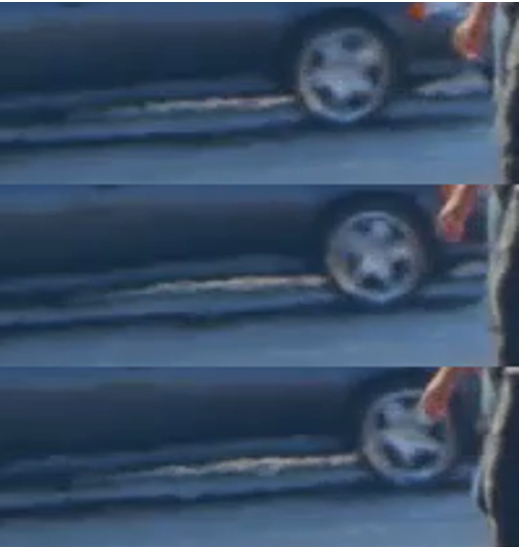}}
    \hfill
  \subfloat[Processed by \cite{anantrasirichai2022atmospheric}]{%
    \includegraphics[width=0.33\linewidth, height=0.3465\linewidth]{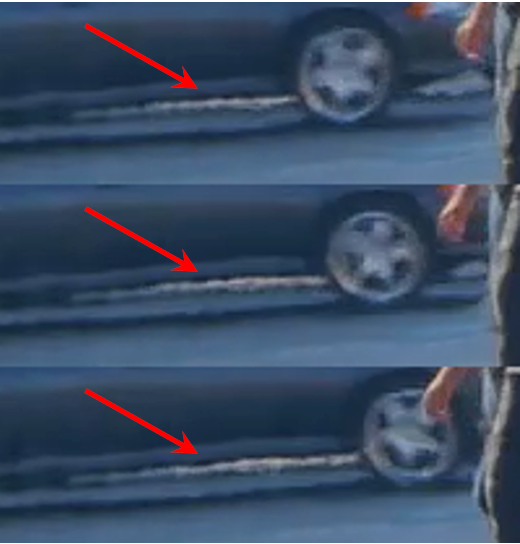}}
   \hfill
  \subfloat[Ours (tilt-removal only)]{%
    \includegraphics[width=0.33\linewidth, height=0.3465\linewidth]{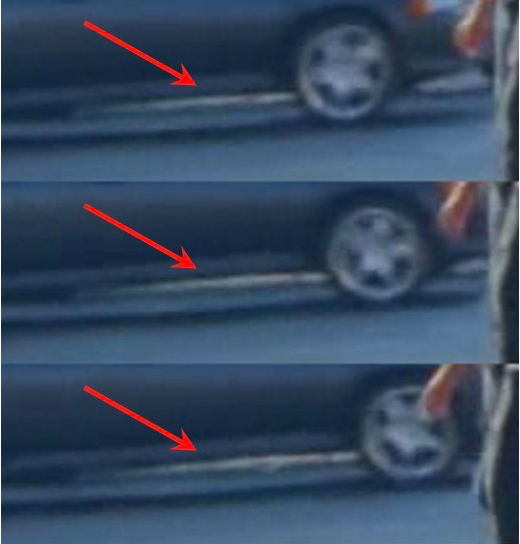}}
  \caption{Application of our tilt-removal module on real-world data from \cite{anantrasirichai2022atmospheric}. Images on the top are full-resolution images; below are the same patches in 3 frames starting from them.}
  \label{fig:tilt_street}
\end{figure*}

\paragraph{Compare with the simulation method used in \cite{anantrasirichai2022atmospheric}} Recent learning-based extension \cite{anantrasirichai2022atmospheric} of CLEAR \cite{Anantrasirichai2013} employed nine predefined PSF of atmospheric turbulence introduced in \cite{Hirsch2010} to simulate the degradation. Those PSFs are not physically grounded and lack diversity, so the restoration performance on real-world data is not ideal. Since their trained model was not released, we could not train it on our dataset to demonstrate the potential improvement. Despite this, we can compare the output from our tilt-removal module with the result of \cite{anantrasirichai2022atmospheric} on their real-world videos. As shown in Fig. \ref{fig:tilt_street}, our lightweight tilt-removal module outperforms a more complex model, which suggests the generalization capability is from our synthetic dataset.

\subsection{Comparison With Conventional Methods}
We make a comparison with two state-of-the-art conventional turbulence mitigation methods \cite{Anantrasirichai2013, mao_tci}. Due to the slow processing speed of the conventional methods, we only compare the methods on a random subset of 100 scenes from our synthetic static dataset. The comparison of the 100 scenes is shown in Table \ref{table:comparison}. Conventional methods could still work because our synthetic data has similar properties compared to real-world turbulence-degraded images. For example, the lucky effect can be synthesized with this simulator \cite{mao_P2S}. It is worth mentioning that we used 50 frames as input for conventional methods and 12/15 for learning-based methods.

\begin{table}[ht]
  \caption{Quantitative comparison with conventional turbulence mitigation methods on the 100 scenes dataset. The first two rows are conventional methods.}
  \label{table:comparison}
  \centering
  \setlength{\aboverulesep}{0.2pt}
\setlength{\belowrulesep}{0.2pt}
  \resizebox{0.48\textwidth}{!}{
  \begin{tabular}{lcccc}
    \toprule[1pt]
    Methods  & PSNR & SSIM & CW-SSIM & LPIPS($\downarrow$)\\
    \midrule
     Mao et al. \cite{mao_tci}    & 27.4154 & 0.8174 & 0.9262 & 0.1998\\
     CLEAR \cite{Anantrasirichai2013}   & 26.9969 & 0.8295 & 0.9293 & 0.1986\\
    \midrule
    BasicVSR++ \cite{chan2022generalization}  & 28.0811 & 0.8410 & 0.9368  & 0.1852 \\
     TSRWGAN \cite{Jin2021NatureMI}   & 25.8139 & 0.7977 & 0.9006  & 0.2315\\
     VRT \cite{liang2022vrt}    & 28.3454 & 0.8512 & 0.9411 & 0.1727\\
     TMT-12f [ours]   & \textbf{28.4543} & \textbf{0.8539} & \textbf{0.9448} & \textbf{0.1640} \\
    \bottomrule[1pt]
  \end{tabular}}
\end{table}

Additionally, we compare our proposed method with \cite{Anantrasirichai2013} and \cite{mao_tci} on real-world images. A qualitative comparison is presented in Fig. \ref{fig:Conventional_clear}. CLEAR \cite{Anantrasirichai2013} used a subsample mechanism to select high-quality images among all 75 frames of the input sequence, then used a complex wavelet-based algorithm to fuse selected frames. After those operations, the output is still subjected to explicit denoising, while our method blindly uses the first 12 frames as input, and the noise is mostly compressed. Mao et al. \cite{mao_tci} is based on the lucky fusion algorithm, a popular framework in conventional turbulence mitigation tasks. It has an explicit denoising stage after the fusion, making the result more blurry. It can be seen that our method produced more natural and high-contrast reconstruction without further loss of details.

\section{Conclusion}
In this paper, we presented a complete and generalizable solution for a multi-frame blind atmospheric turbulence mitigation problem. We refined a physics-based simulation in \cite{mao_P2S} by approximating the spatially varying field using a wide sense stationary field. We further proposed an efficient transformer-based network to restore image sequences degraded by turbulence, and our method can adapt to various scenes and turbulence conditions. Our model surpassed the state-of-the-art models designed for general video restoration tasks. Compared to conventional methods of turbulence mitigation, the proposed method has superior performance in visual quality, speed, and robustness.

\bibliographystyle{IEEEtran}
\bibliography{IEEEabrv, ref}

\end{document}